\documentclass[letterpaper,12pt, abstracton]{scrartcl}

\usepackage{marvosym}
\usepackage[T1]{fontenc}
\usepackage{color,hyperref}
\usepackage{enumitem}
\usepackage{amssymb}
\usepackage{amsmath}
\usepackage{wasysym}
\usepackage{pifont}
\usepackage{makecell}
\usepackage{subcaption}
\usepackage{booktabs}
\usepackage{Sweave}

\newcommand\Rey{\mbox{\textit{Re}}}  
\newcommand\Wi{\mbox{\textit{Wi}}}  
\date{}

\begin{document}

\title{Is a UCM fluid flow near a stationary point always singular?}
\author{Igor Mackarov \\ \href{mailto:Mackarov@gmx.net}{{\small Mackarov@gmx.net}}}

\maketitle
\begin{abstract}
Frequently observed divergence of numerical solutions to benchmark flows of the UCM viscoelastic fluid is a known and widely discussed issue. Some authors consider such singularities ``invincible''. The article argues this position, to which end it considers two typical flows with a  stagnation point, often a place of the flows' singularity: counterflows and a flow spread over a wall. For the  counterflows numerical and asymptotic analytical solutions are presented. Both kinds of  flows turn out regular in the stagnation points, in particular, for high Weissenberg numbers. A good accordance is demonstrated between the analytical and numerical results.


\end{abstract}

\section{Introduction}
\label{Intro}
 Benchmark flows of viscoelastic fluids such as a flow near a flat wall or past a cylinder or counterflows within cross slots  often pose a lot of complicated features and challenges in their description \cite{Nonlinearity}. Among the problems there are
\begin{itemize}
\item formation of various forms of secondary flows or vortexes often leading to purely elastic instabilities \cite{Pathak,LiXi,Groisman97, Groisman98, Benchmark},
\item spontaneous onsets of flow asymmetries near stationary points \cite{Poole, Rocha, MackarovBif, Mackarov2014,Couette},
\item reversals of flows \cite{Renardy2006,Mackarov2011,Mackarov64} (shown here as well),
\item emergence of singularities in a flow, in particular, significant
strains in the vicinity of a stagnation point yielding local zones of extremely high
\end{itemize}
 \section*{}
\noindent
 stresses --- sometimes in the form of boundary layers and birefringent strands \cite{Thomases,Becherer,Objections,Vajravelu, Wapperom, Renardy2000}.

Regarding the last item, it is common to observe appearance of  singularities in special points of an important class of \textit{stagnation flows} \cite{LiXi, Pipea2009}. With respect to the UCM model, some authors tend to consider such emergence ``near special points where the velocity vanishes
 --- even though the geometry is not singular'' \cite{Objections} --- as an intrinsic value of this model. At times doubts are raised about general possibility to satisfy the UCM rheological law and momentum equation simultaneously \cite{VanGorder}.

Here we are going to dispute inevitability of the stresses singularity in stationary and stagnation points. For this we will consider two problems, \textit{viz.} the counterflows within the cross slots and the spread along a wall (shortly) --- Sections \ref{sec:counterflows}, and \ref{sec:spread}. We will present regular analytical asymptotic solution to the counterflows near the stationary point. Further, numerical solution to the counterflows (in particular, for high Weissenberg numbers) will be given and thoroughly compared to the analytical results. We will then discuss the used numerical procedure  (Section \ref{sec:NumDetails}). Finally, the main conclusions from the research done will be presented in the context of existing results.

 \begin{figure*}[t]
  \centering
  \includegraphics[width=0.5\textwidth]{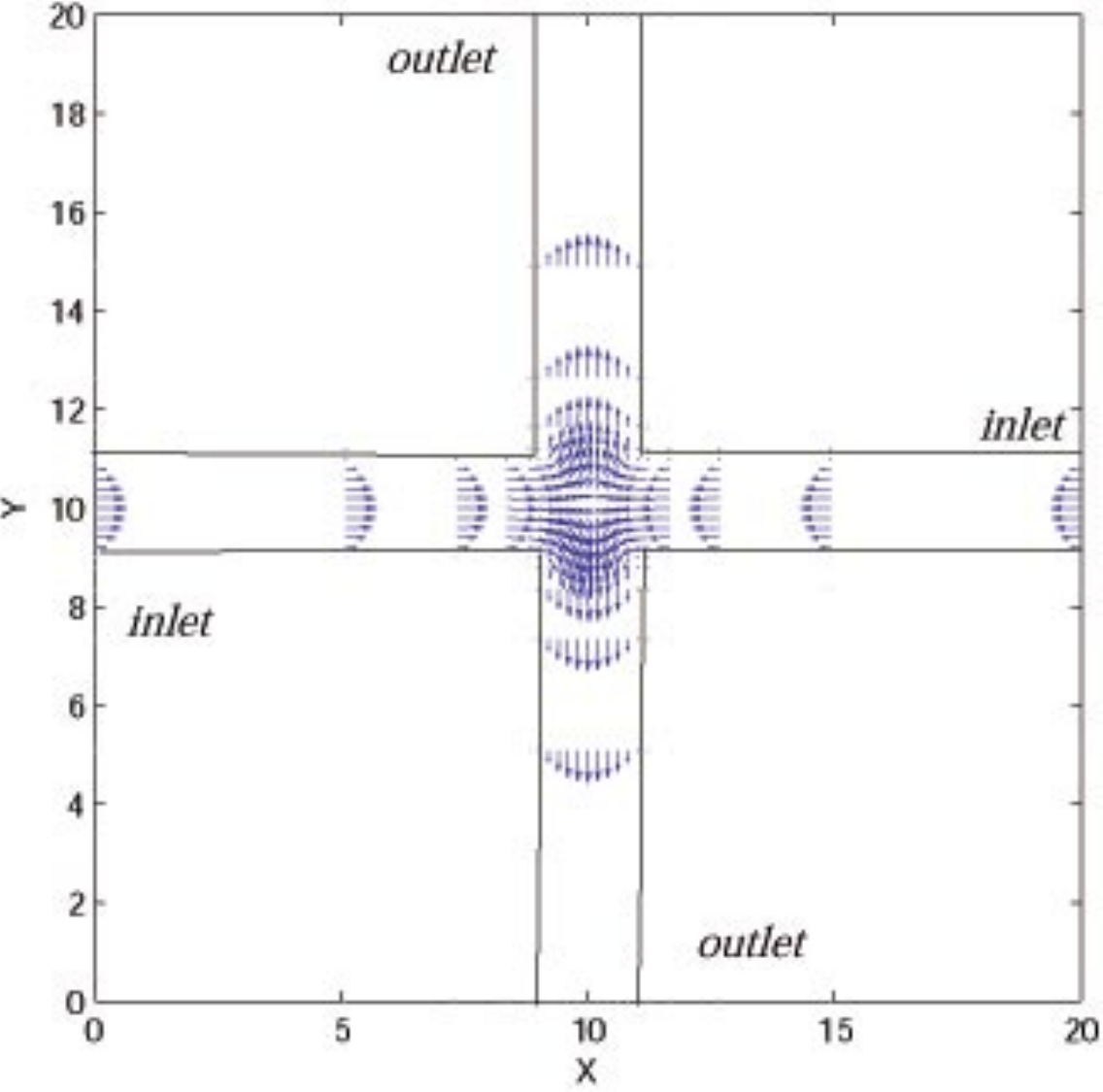}
  \caption{General layout of the flows considered.} 
   \label{fig:layout}
\end{figure*}

\section{Counterflows}
\label{sec:counterflows}
 \subsection{The problem statement}
We thus consider two flows moving along a pair of horizontal slots towards each other and spreading along vertical ones (Fig.~\ref{fig:layout}).

In terms of the dimensionless variables normalized on the problem natural scales (asymptotic stationary inlet pressure ${p_{inlet}}$, the fluid density $\rho$, the velocity scale $\sqrt {{\raise0.7ex\hbox{${{p_{inlet}}}$} \!\mathord{\left/
			{\vphantom {{{p_{inlet}}} \rho }}\right.\kern-\nulldelimiterspace}
		\!\lower0.7ex\hbox{$\rho $}}} $~, and a horizontal slot semi-width) the UCM constitutive equation is written as

 \begin{equation}
 \frac{{{D_O}\,{\sigma _{ij}}}}{{{D_O}\,t}}  + \frac{1}{{\Wi}}{\sigma _{ij}} =  \,\frac{1}{2}\frac{1}{{\Wi \Rey}}\left( {\frac{{\partial {v_i}}}{{\partial {x_j}}} + \frac{{\partial {v_j}}}{{\partial {x_i}}}} \right),\,\,\,\,\,\,\,\,\,\,\,\,i,\,j = 1,2.
 \label{UCM}
 \end{equation}

 \noindent Here the Oldroyd derivative of the stress tensor (henceforth, the Einstein notation is used, which means summation on the repeated indexes) is

 \begin{equation}
 \frac{{{D_O}\,{\sigma _{ij}}}}{{{D_O}\,t}} \equiv \frac{{\partial {\sigma _{ij}}^{}}}{{\partial \,t}} + {v_k}{\frac{{\partial \,{\sigma _{ij}}}}{{\partial \,{x_k}}}_{}} - \frac{{\partial \,{v_i}}}{{\partial \,{x_k}}}{\sigma _{kj}}^{} - \frac{{\partial \,{v_j}}}{{\partial \,{x_k}}}{\sigma _{ik}},\,\,\,\,\,\,\,\,\,\,\,\,i,\,j,\,k = 1,2.
 \label{Jauman}
 \end{equation}

 \noindent The flows are also governed by the equations of momentum
 \begin{equation}
 \frac{{\partial \,{v_i}}}{{\partial \,{t}}} + {v_j}\;\frac{{\partial \,{v_i}}}{{\partial \,{x_j}}} =  - \frac{{\partial p}}{{\partial \,{x_i}}} + \frac{{\partial {\kern 1pt} {\sigma _{ij}}}}{{\partial \,{x_j}}}\,,\,\,\,\,\,\,\,\,\,\,\,i,\,j = 1,2
 \label{momentum}
 \end{equation}

 \noindent and continuity
 \begin{equation}
 \frac{{\partial \,{v_i}}}{{\partial \,{x_i}}} = 0,\,\,\,\,\,\,\,\,\,\,\,\,\,i = 1,2.
 \label{continuity}
 \end{equation}

 As \textit{initial conditions} zero values are set for all the dependent variables.

 \textit{Boundary conditions} consist in no-slip constraints at the walls \begin{math} {u_{wall}  = v_{wall}  = 0} \end{math},  setting the outlet pressure to 0, and an increase of the inlet pressure from 0 up to 1 on  a smooth law  to reach a quasi-steady flow. Among a few laws tried, which eventually resulted in rather close solutions, following expression was used to obtain the results presented in this paper:
\begin{equation}
 {p_{inlet}(t)={\alpha t} \mathord{\left/
   {\vphantom {{ kt} {(1 +\alpha t)}}} \right.
   \kern-\nulldelimiterspace} {(1 + \alpha t)}}.
\end{equation}
Varying $\alpha$ from 1 up to 10 did not bring about any remarkable changes in the solutions either, and the results below are all gotten with $\alpha=1$ (cf. \cite{Mackarov2011}). When $t =25$, the inlet pressure time derivative is less than 0.15\% of its initial value. So later on, the flow conditions are considered near-stationary.

\subsection{Asymptotic solution to the symmetric stationary flow}
\label{sec:asymp:counterflows}
Expressions for the velocities, stresses and pressure of such UCM fluid counterflows near the central point ($x=y=0$)\footnote{For better presentation, the coordinates
in Fig. \ref{fig:layout} and all next figures pertinent to the counterflows and spread are shifted by 10 units so that the stationary points there are $(10,10)$ and $(0,10)$, respectively.} satisfying system (\ref{UCM})--(\ref{continuity}) in the stationary symmetric case \textit{up to the third order on} $x,\  y$ were derived and discussed in \cite{MackarovBif,Mackarov2014}. For the sake of completeness remind them here:
\begin{equation}
u\left( {x,y} \right) = A  x + B  x {y^2} + \frac{1}{3}B  {x^3},
\label{counter.u.symmetric}
\end{equation}
\begin{equation}
v\left( {x,y} \right) =  - A  y - B {x^2}y - \frac{1}{3}B  {y^3},
\label{counter.v.symmetric}
\end{equation}
\vspace{0.05cm}
\begin{equation}
{\sigma _{xx}}\left( {x,y} \right){\rm{ = }}{\Sigma _x} + \alpha {x^2} + \beta{y^2},
\label{counter.sigma.xx}
\end{equation}
\vspace{0.1cm}
\begin{equation}{\sigma _{yy}}\left( {x,y} \right){\rm{ = }}{\Sigma _y} + \delta  {x^2} + \varepsilon  {y^2},
\label{counter.sigma.yy}
\end{equation}
\vspace{0.1cm}
\begin{equation}
\label{counter.sigma.xy}
{{\rm{\sigma }}_{xy}}\left( {x,y} \right){\rm{ = }}\kappa  x y,
\end{equation}
\vspace{0.1cm}
\begin{equation}
\label{counter.pressure}
p(x,y) = {P_0} + {P_x}{x^2} + {P_y}{y^2},
\end{equation}
\\
with coefficients
\begin{equation}
{\Sigma _x}(\textit{A}) =  - \frac{A}{{\Rey\left( {2A \Wi - 1} \right)}},
\label{counter.Sigma.x}
\end{equation}
\begin{equation}
\label{counter.Sigma.y}
{\Sigma _y}(\textit{A})\equiv - {\Sigma _x}(-\textit{A}) =  - \frac{A}{{\Rey\left( {2A \Wi + 1} \right)}},
\end{equation}
\begin{equation}
\alpha (\textit{A},\textit{B}) =  - \frac{B}{{\Rey\left( {2A\Wi - 1} \right)}},
\label{alpha}
\end{equation}
\begin{equation}
\beta(\textit{A},\textit{B}) = \frac{B}{{\Rey\left( {2A\Wi - 1} \right)\left( {4A\Wi - 1} \right)}},
\label{beta}
\end{equation}
\begin{equation}
\varepsilon ({\textit{A}},{\textit{B}}) \equiv \alpha(-{\textit{A}},-{\textit{B}}) =  - \frac{B}{{\Rey\left( {2A\Wi + 1} \right)}},
\end{equation}
\begin{equation}
\delta ({\textit{A}},{\textit{B}}) \equiv \beta (-{\textit{A}},-{\textit{B}})
 =  - \frac{B}{{\Rey\left( {2A\Wi + 1} \right)\left( {4A\Wi + 1} \right)}},
\end{equation}
\begin{equation}
\label{counter.kappa}
\kappa(\textit{A},\textit{B})  = 4\frac{{A B \Wi}}{{Re\left( {{{\left( {2AWi} \right)}^2} - 1} \right)}},
\end{equation}
\begin{equation}
{P_x}(A,B) = -\frac{A^2}{2} + \frac{{B}}{{\Rey \left( {1 - {{(2A\Wi)}^2}}\right )}},
\label{counter.p.x}
\end{equation}
\begin{equation}
{P_y}(A,B) \equiv {P_x}( - A, - B) = -\frac{A^2}{2} - \frac{{B}}{{\Rey \left( {1 - {{(2A\Wi)}^2}} \right)}}.
\label{counter.p.y}
\end{equation}
Herewith, we assume the validity of the variables power expansions, i.e., existence of the velocities' fourth derivatives and the third derivatives of the rest of the variables. It should be noticed that the flow near the center was found to weakly depend on the global flow conditions through the coefficients \textit{A} and \textit{B}~ \cite{MackarovBif}~, whose values must be determined by merging the asymptotic and global solutions.

Thus, according to Eqs.~(\ref{counter.u.symmetric})--(\ref{counter.p.y}), a stationary point between the counterflows \textit{is not obliged }to be singular --- contrary to the assertions of \cite{Objections} with respect to the UCM model for any flow.

Evidently, regular power expansions cannot involve values $A =  \pm 1/2Wi,\;\pm 1/4Wi$.
 In practice, as a rule, such values are not the case, which will be exhibited  below while discussing the numerical solutions and providing an example of a merge between the numerical and global solutions.

\subsection{Principal numerical results}
\label{sec:princ:counterflows}

\begin{figure*}[t]
        \centering
        \begin{subfigure}[]{0.3\textwidth}
                \includegraphics[width=1\textwidth]{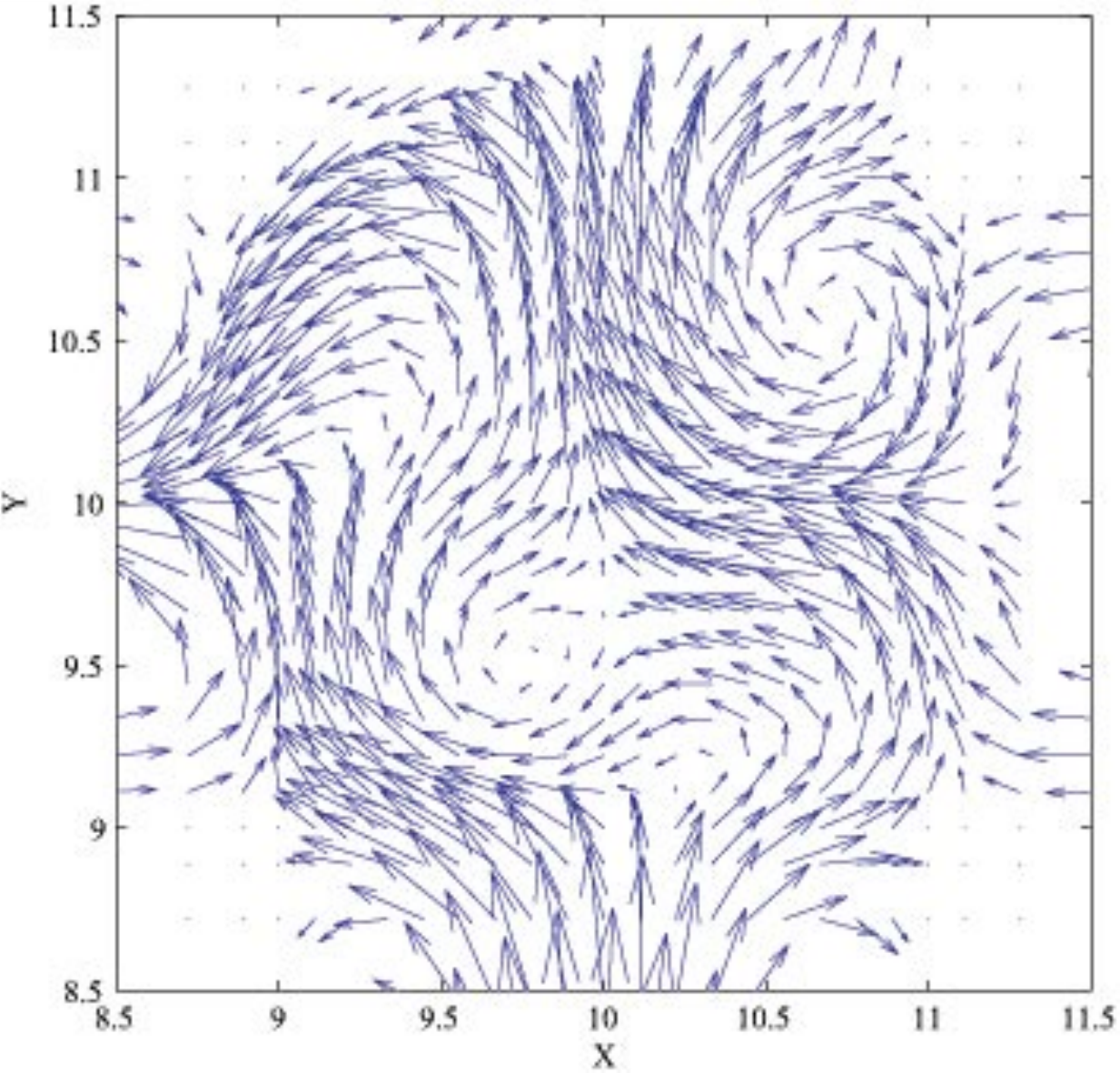}
                \caption{\scriptsize{$\Delta t = 10^{-3}$,\\ $ \Delta = 0.10$.}}
        \end{subfigure}
        \begin{subfigure}[]{0.3\textwidth}
                           \includegraphics[width=1\textwidth]{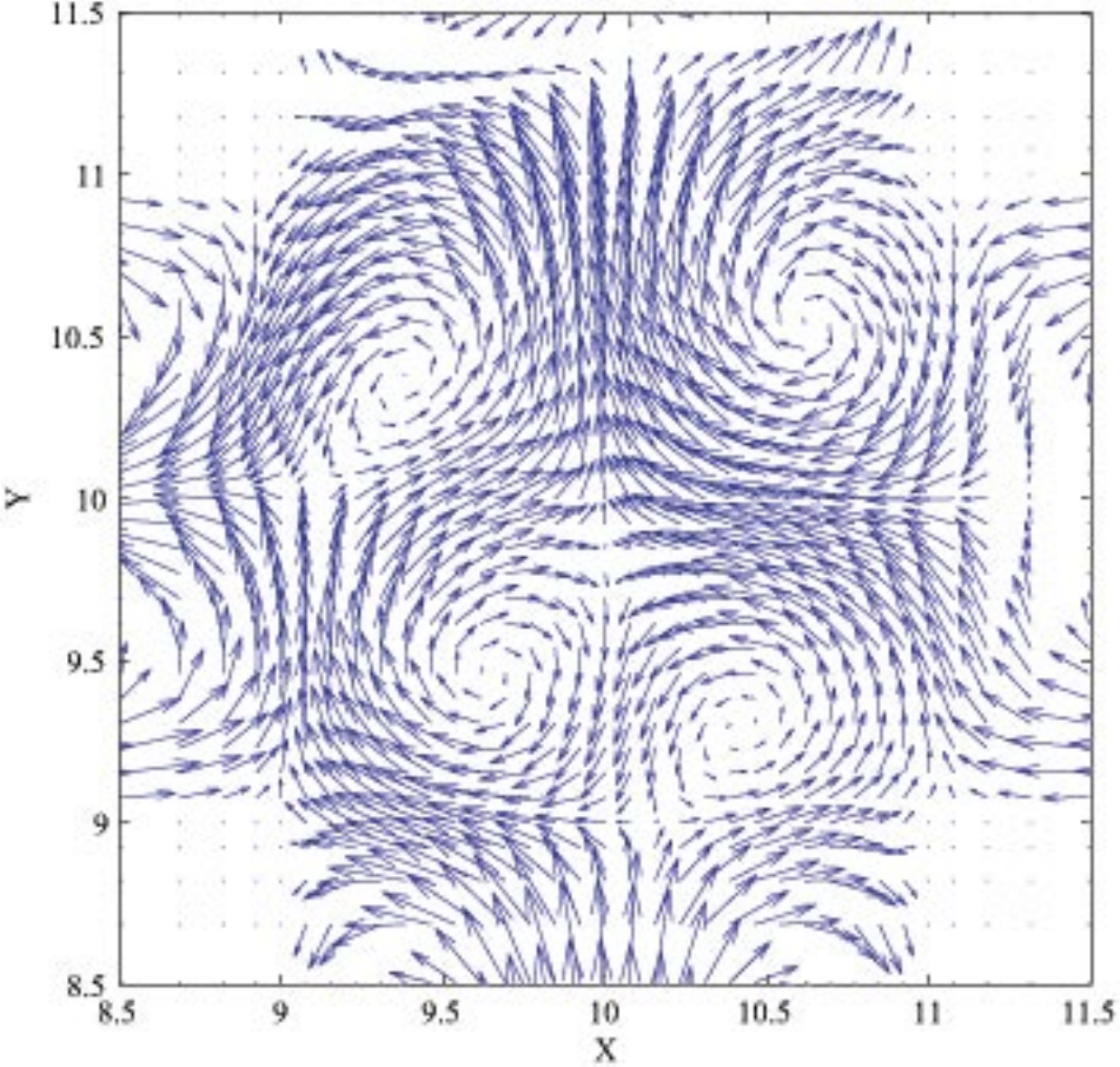}
                           \caption{\scriptsize{$\Delta t =5\cdot{10^{-4}}$,\\  $\Delta = 0.071$.}}
          \end{subfigure}

        \begin{subfigure}[]{0.3\textwidth}
                           \includegraphics[width=1\textwidth]{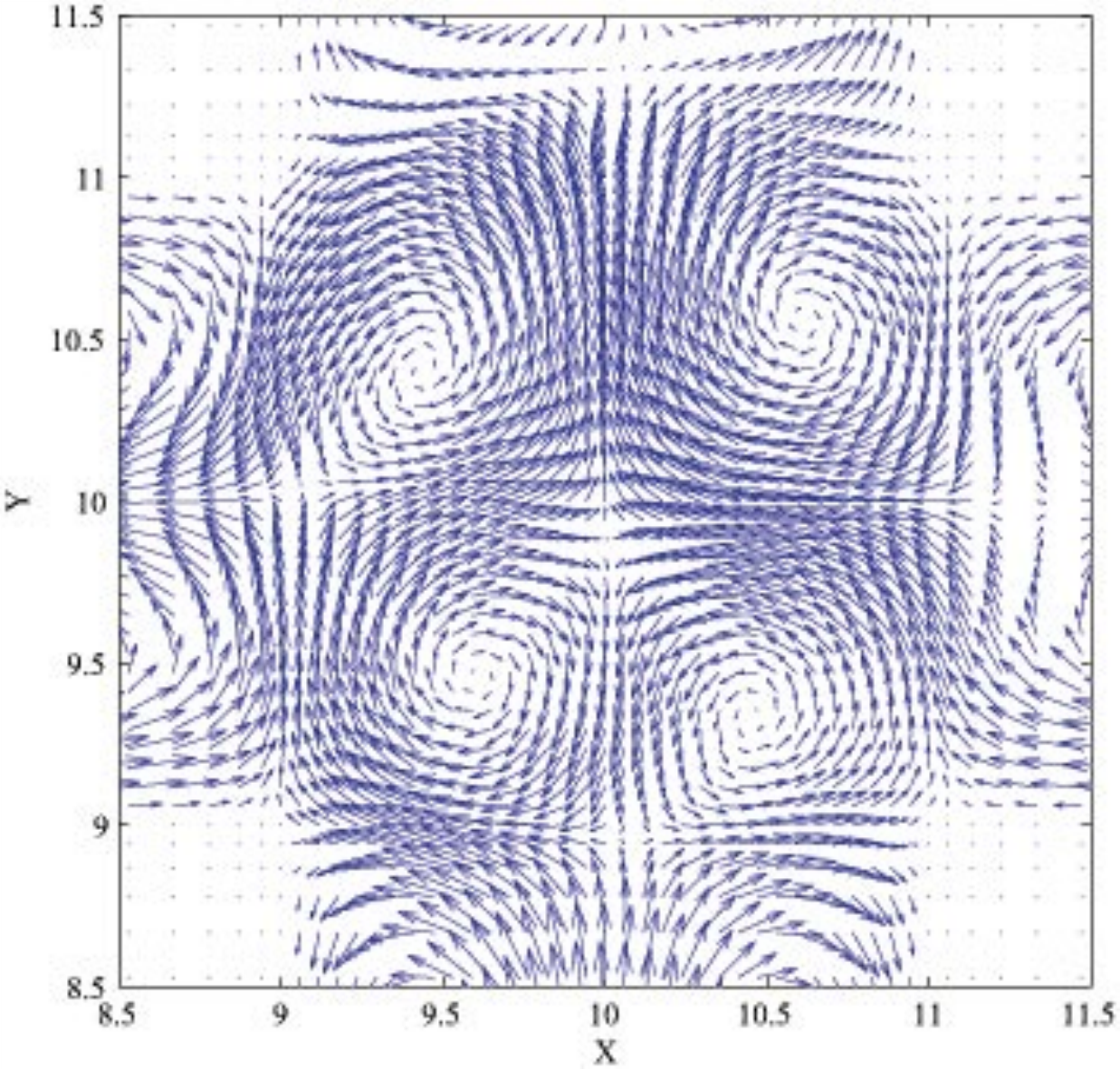}
                           \caption{\scriptsize{$\Delta t =10^{-4}$,\\  $\Delta = 0.056$.}}
       \end{subfigure}
      \begin{subfigure}[]{0.3\textwidth}
                                  \includegraphics[width=1\textwidth]{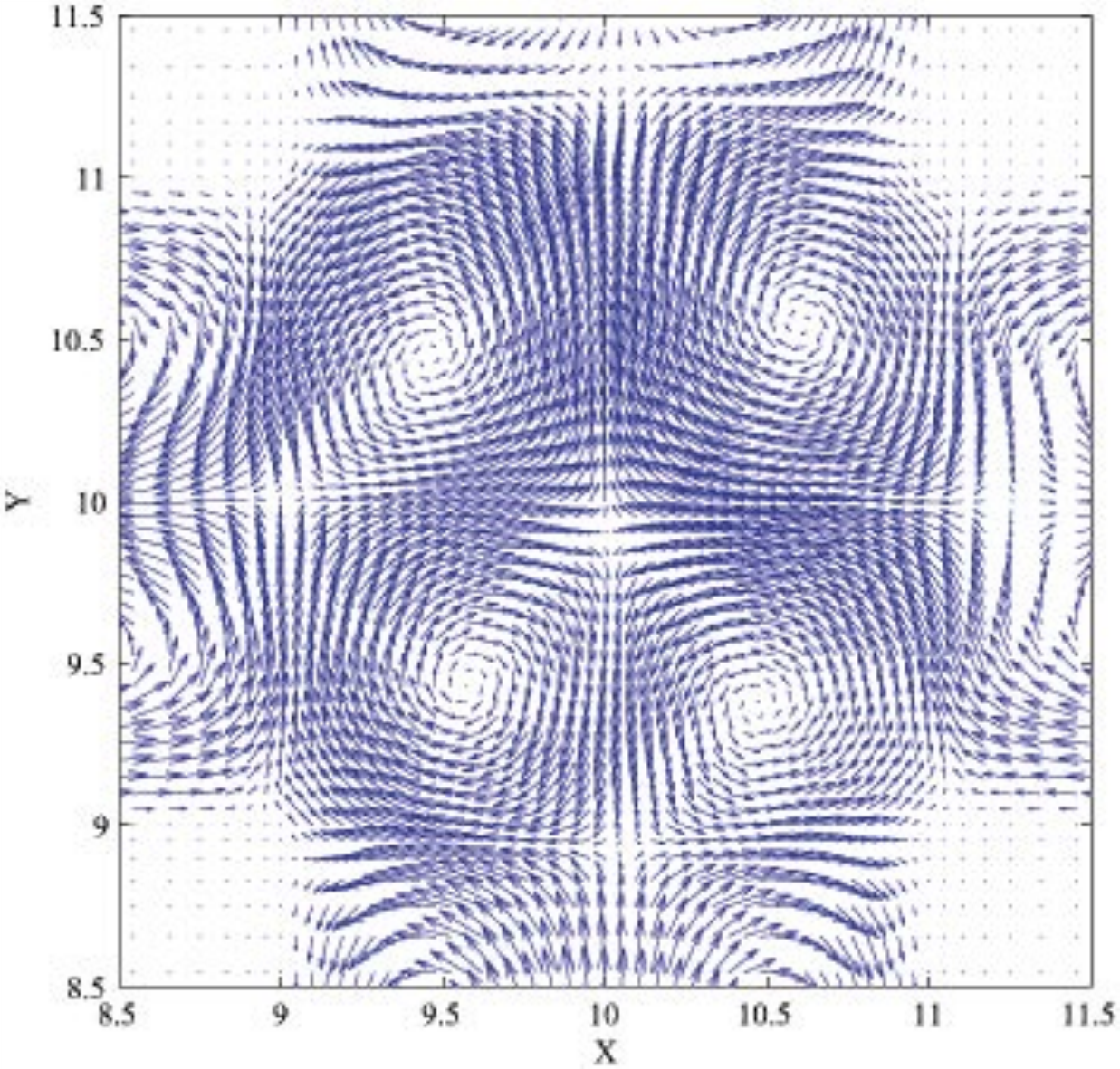}
                                  \caption{\scriptsize{$\Delta t =10^{-4}$, \\  $\Delta = 0.046$.}}
      \end{subfigure}
                 \caption{An instance of the counterflows  given by the numerical simulation on different meshes; $t=6.3$, $\Rey=0.01$, $\Wi=100$. Specified are the steps on time and space: $\Delta t$ and $ \Delta$.}
                      \label{fig:VortexCounterPattern}
\end{figure*}

\begin{figure*}[t]
        \centering
        \begin{subfigure}[]{0.3\textwidth}
                \includegraphics[width=1\textwidth]{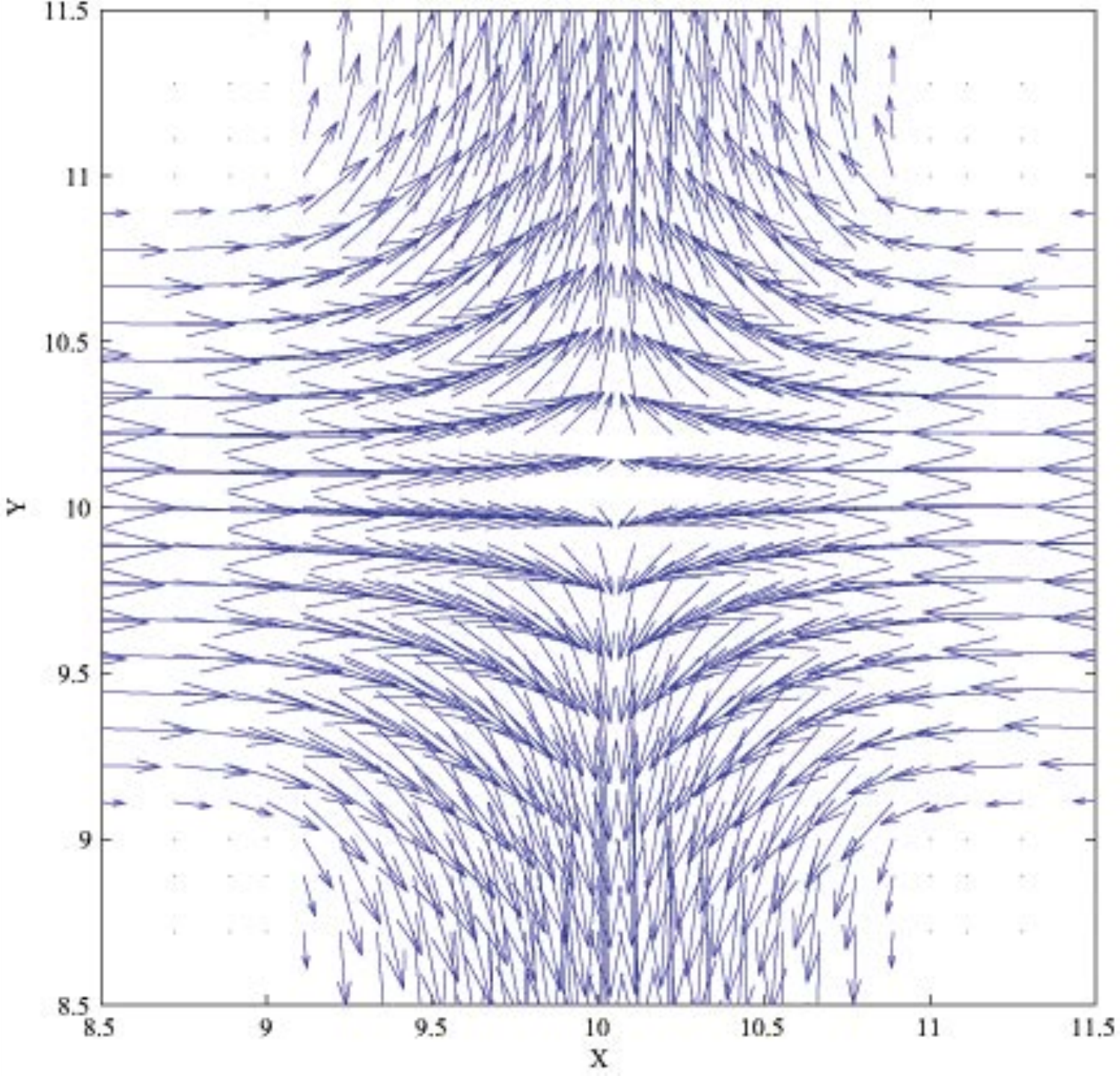}
                \caption{\scriptsize{$\Delta t = 10^{-3}$,\\ $ \Delta = 0.10$.}}
        \end{subfigure} ~~
        \begin{subfigure}[]{0.3\textwidth}
                           \includegraphics[width=1\textwidth]{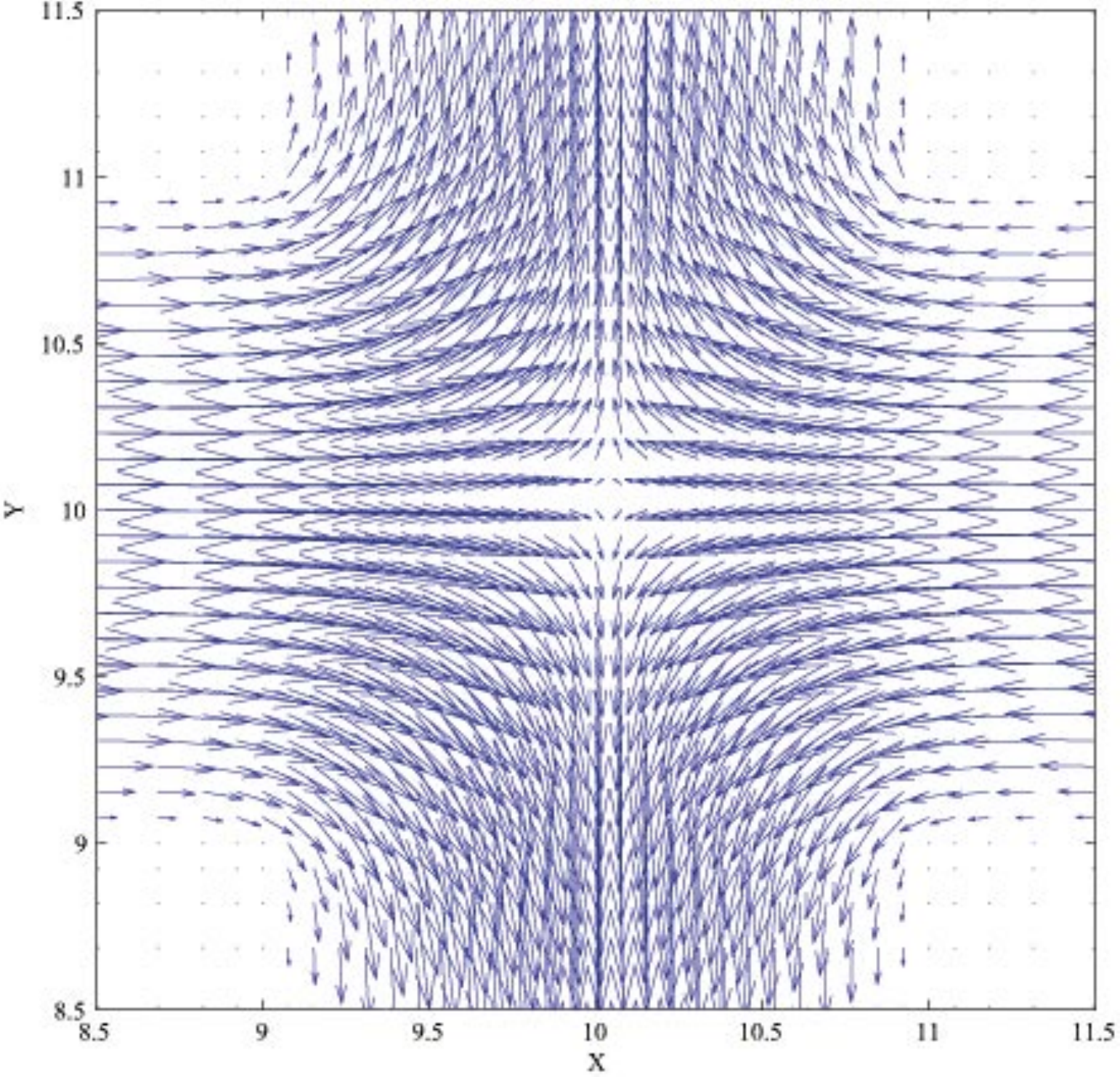}
                           \caption{\scriptsize{$\Delta t =5\cdot{10^{-4}}$,\\  $\Delta = 0.071$.}}
          \end{subfigure}

        \begin{subfigure}[]{0.3\textwidth}
                           \includegraphics[width=1\textwidth]{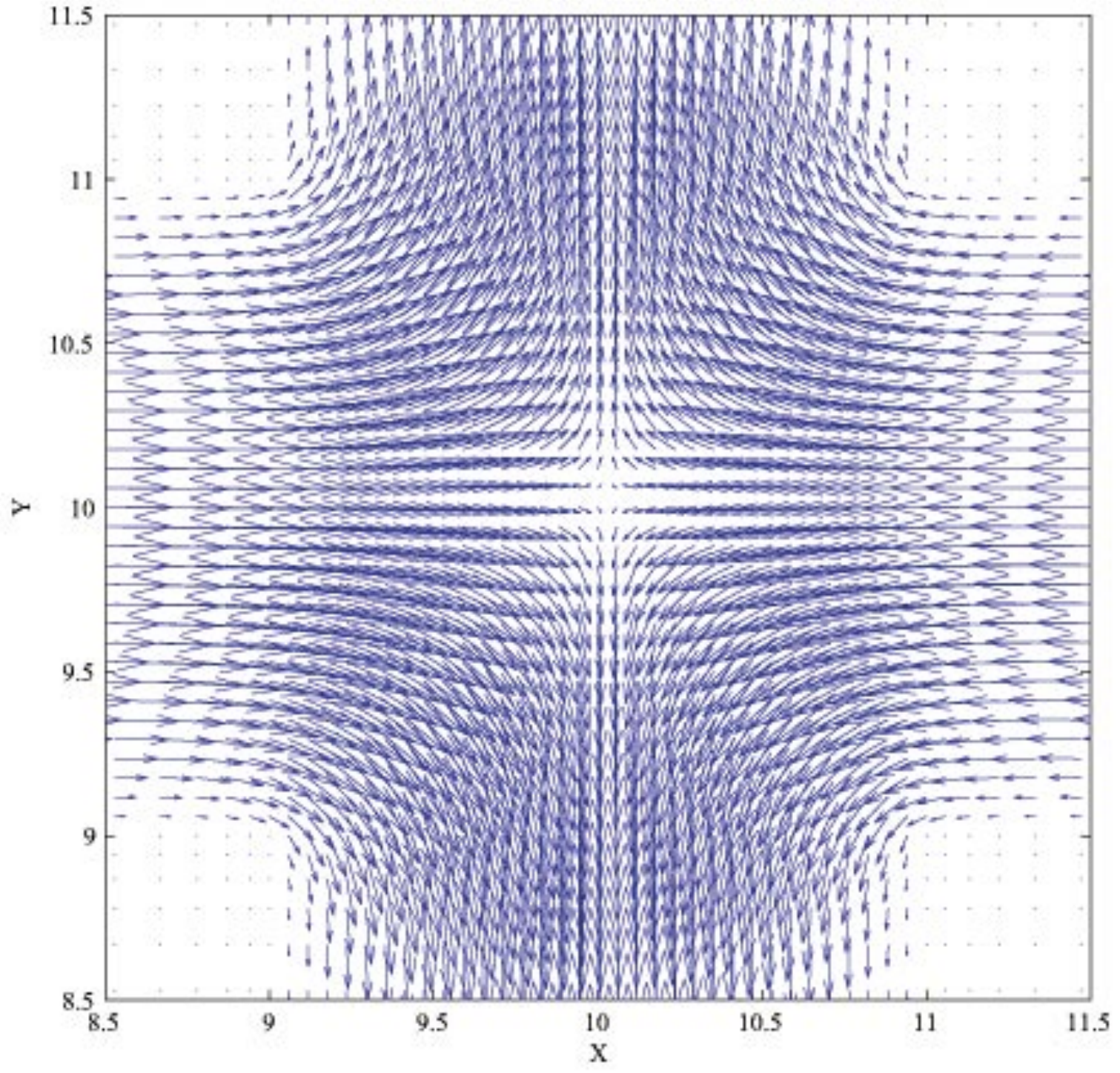}
                           \caption{\scriptsize{$\Delta t =10^{-4}$,\\  $\Delta = 0.056$.}}
       \end{subfigure} ~~
      \begin{subfigure}[]{0.3\textwidth}
                                  \includegraphics[width=1\textwidth]{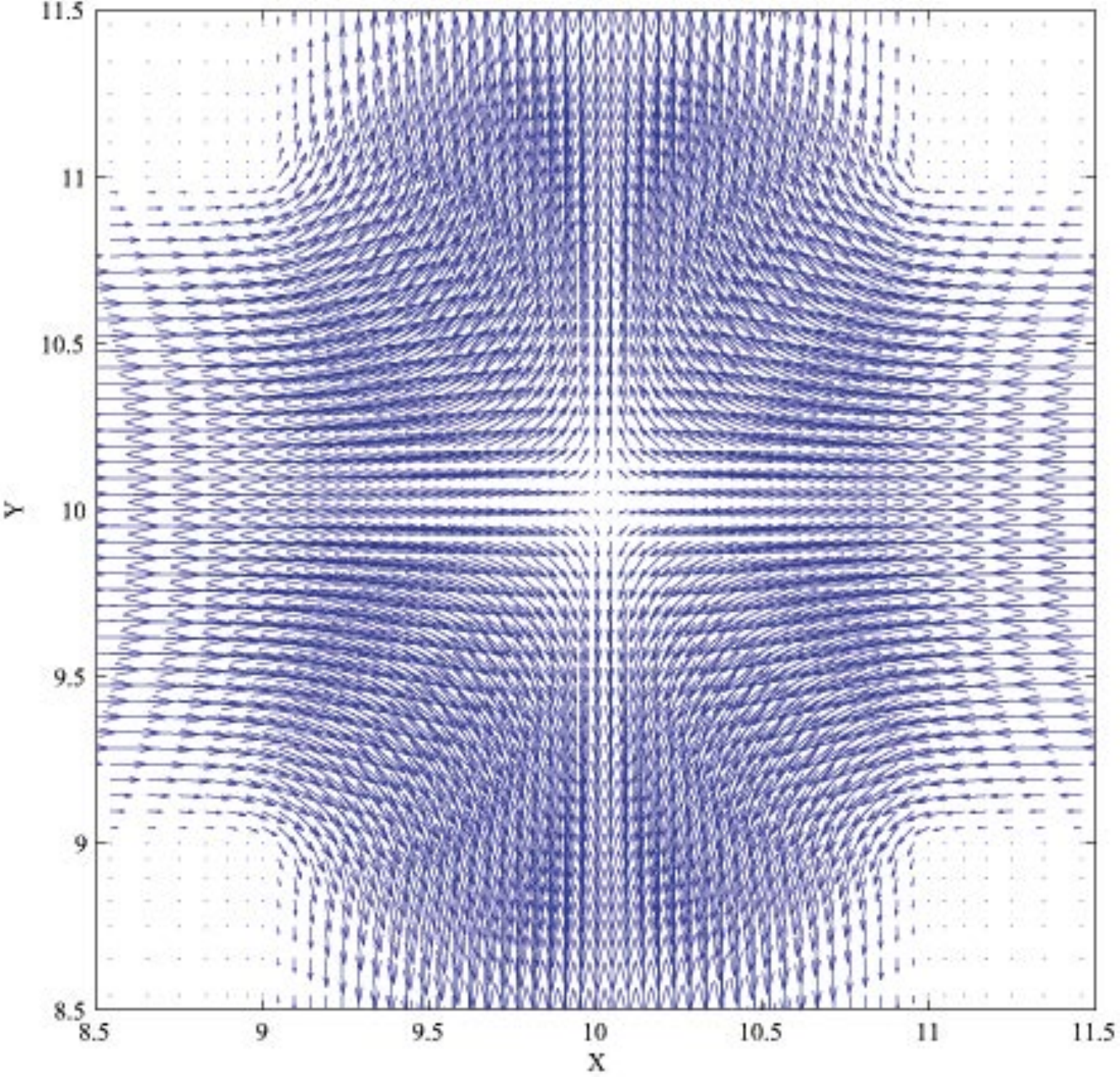}
                                  \caption{\scriptsize{$\Delta t =10^{-4}$, \\  $\Delta = 0.046$.}}
      \end{subfigure}
                 \caption{An instance of the counterflows  given by the numerical simulation on different meshes; $t=7.9$, $\Rey=0.01$, $\Wi=100$.  Specified are the steps on time and space: $\Delta t$ and $ \Delta$.}
                      \label{fig:RegularCounterPattern}
\end{figure*}

As was previously  detected analytically and numerically \cite{Mackarov2011,Mackarov64}, the process of counterflows stabilization is characterized by a periodic formation of vortex-like structures with a simultaneous change of the flows direction. One of such vortexes intensities tops is caught by Fig.
~\ref{fig:VortexCounterPattern}. Some later the counterflows will take a regular structure (Fig.
~\ref{fig:RegularCounterPattern}) until they again change directions with a newly emerged group of vortexes.

The counterflows with moderate Weissenberg numbers usually come to a stable regime after a few of such reversals. However, in highly-elastic cases ($\Wi>10$) periodic flows reversals accompanied by vortexes emergences can last for a very long time.

On four various domain partitions (Figs.~\ref{fig:VortexCounterPattern},\  \ref{fig:RegularCounterPattern})  the counterflows fields obviously look like subsequent refinements of a common flow structure. This is set to show a good numerical convergence --- even though the Weissenberg number is really high!

More strictly and quantitatively the convergence will be discussed in Section~\ref{sec:counter:convergence}.

\subsection{Comparison of the analytical and numerical solutions}

\begin{figure*}[t]
  \centering
  \resizebox{7.5cm}{!}{\includegraphics{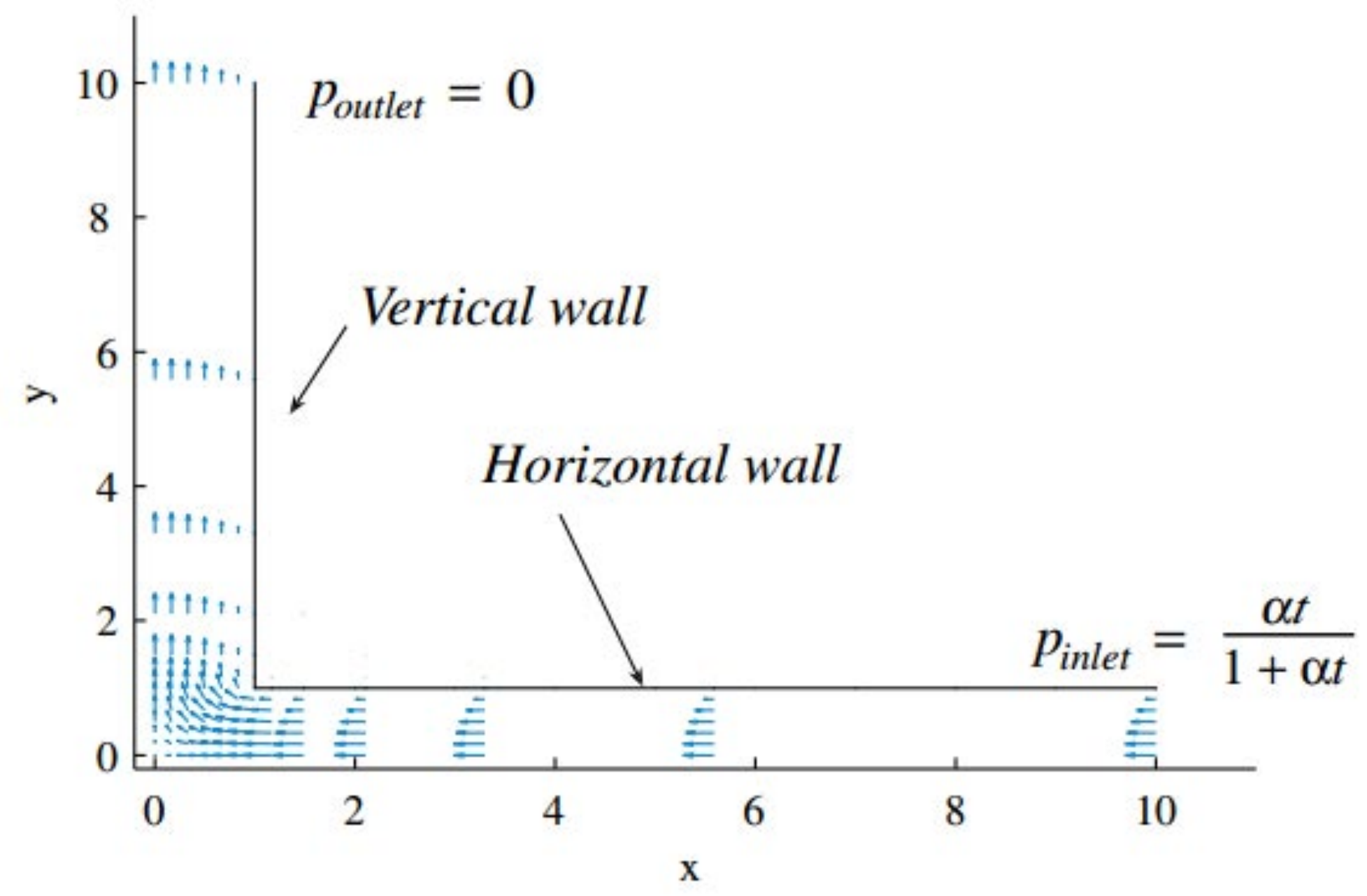}}
  \caption{Layout of counterflows in a one-quarter domain.}
  \label{fig:1q-lay}
\end{figure*}

\begin{figure*}[t]
        \centering
               \begin{subfigure}[t]{0.45\textwidth}
               	\centering
                        \includegraphics[width=1\textwidth]{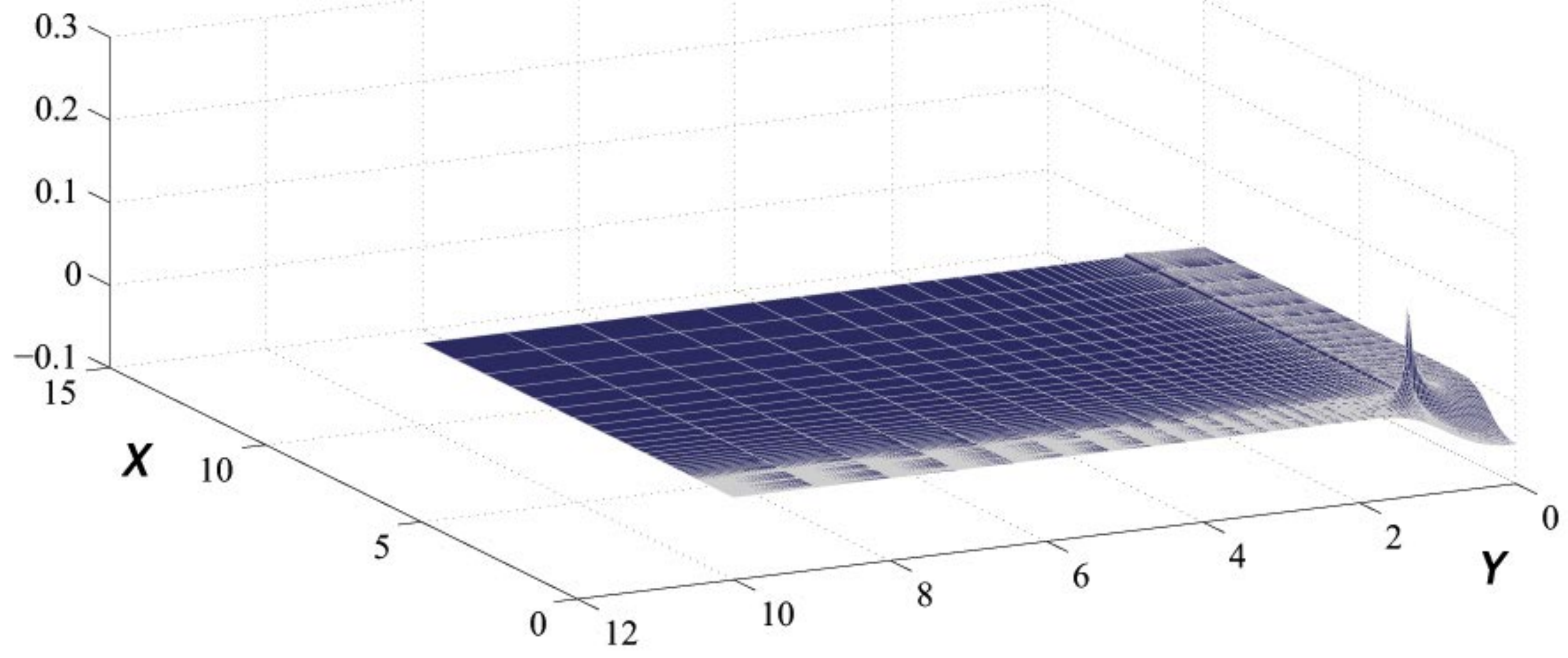}
                        \caption{Normal stress - one quadrant.}
                        \label{fig:StressAnalNum:quarter}
                \end{subfigure} ~~
                \begin{subfigure}[t]{0.45\textwidth}
                	\centering
                        \includegraphics[width=0.75\textwidth]{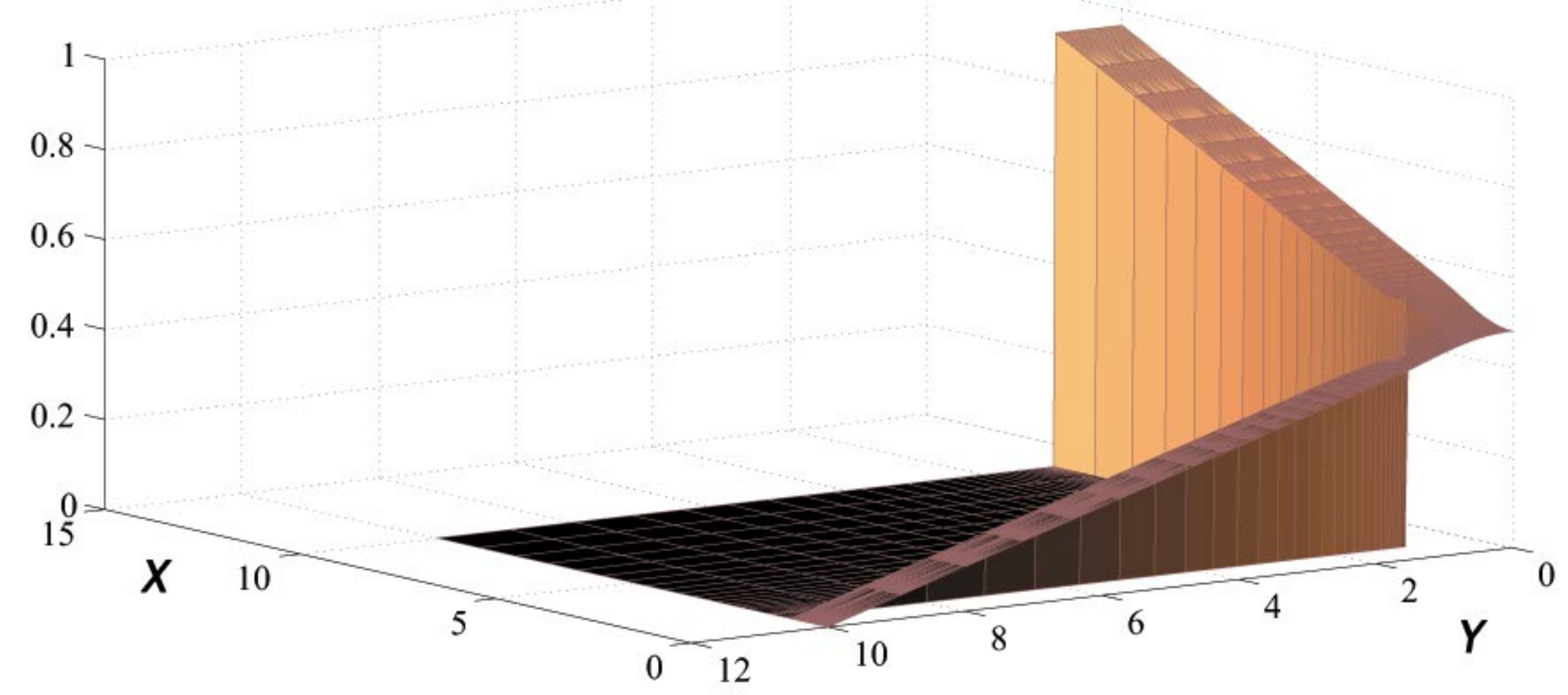}
                        \caption{Pressure - one quadrant.}
                        \label{fig:PressureAnalNum:quarter}
                \end{subfigure}

        \begin{subfigure}[b]{0.45\textwidth}
        	\centering
                \includegraphics[width=0.6\textwidth]{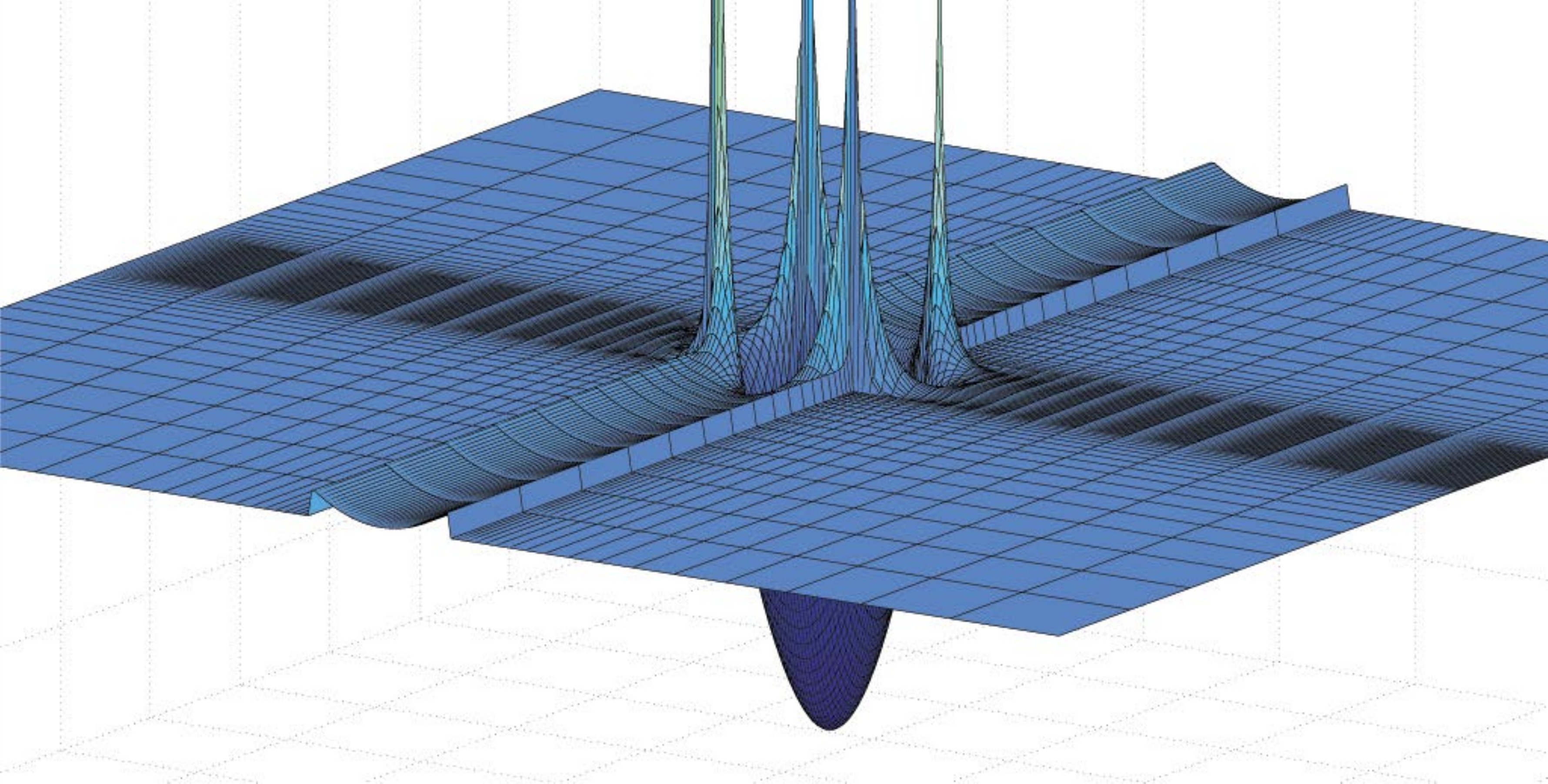}
                \caption{Normal stress - whole domain.}
                \label{fig:StressAnalNum}
        \end{subfigure}
        \begin{subfigure}[b]{0.45\textwidth}
        	\centering
                \includegraphics[width=0.6\textwidth]{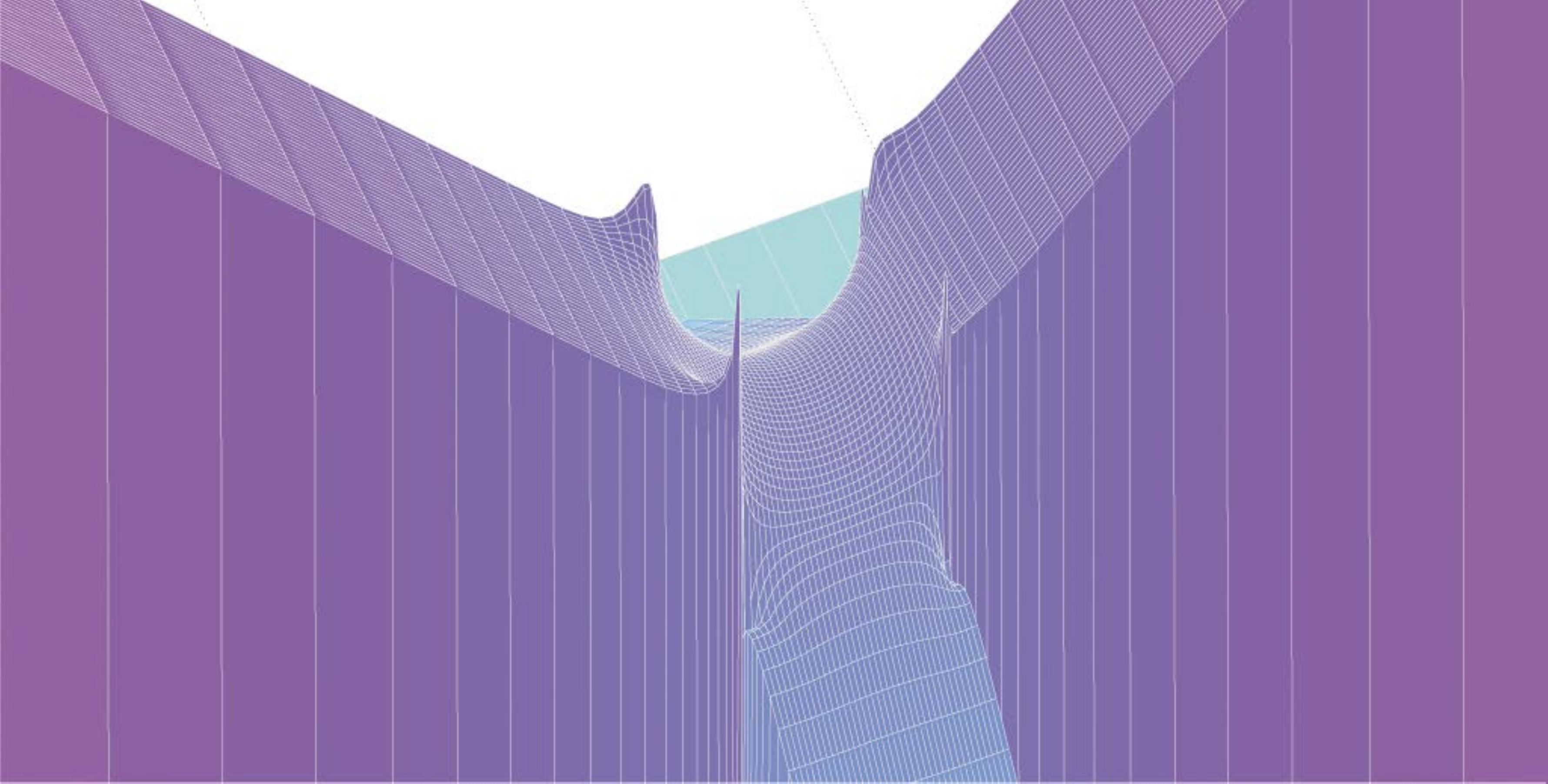}
                \caption{Pressure - whole domain.}
                \label{fig:PressureAnalNum}
        \end{subfigure}
      \caption{Distributions of the flow characteristics for the numerical solutions to the counterflows in a steady regime ($t=30$). \Rey = 0.1 and \Wi = 4. The minimum finite difference step on both size dimensions is 0.03.}
           \label{fig:PressureStressAnalNum}
\end{figure*}

\begin{figure*}[t]
        \centering
        \begin{subfigure}[b]{0.45\textwidth}
                \includegraphics[width=1\textwidth]{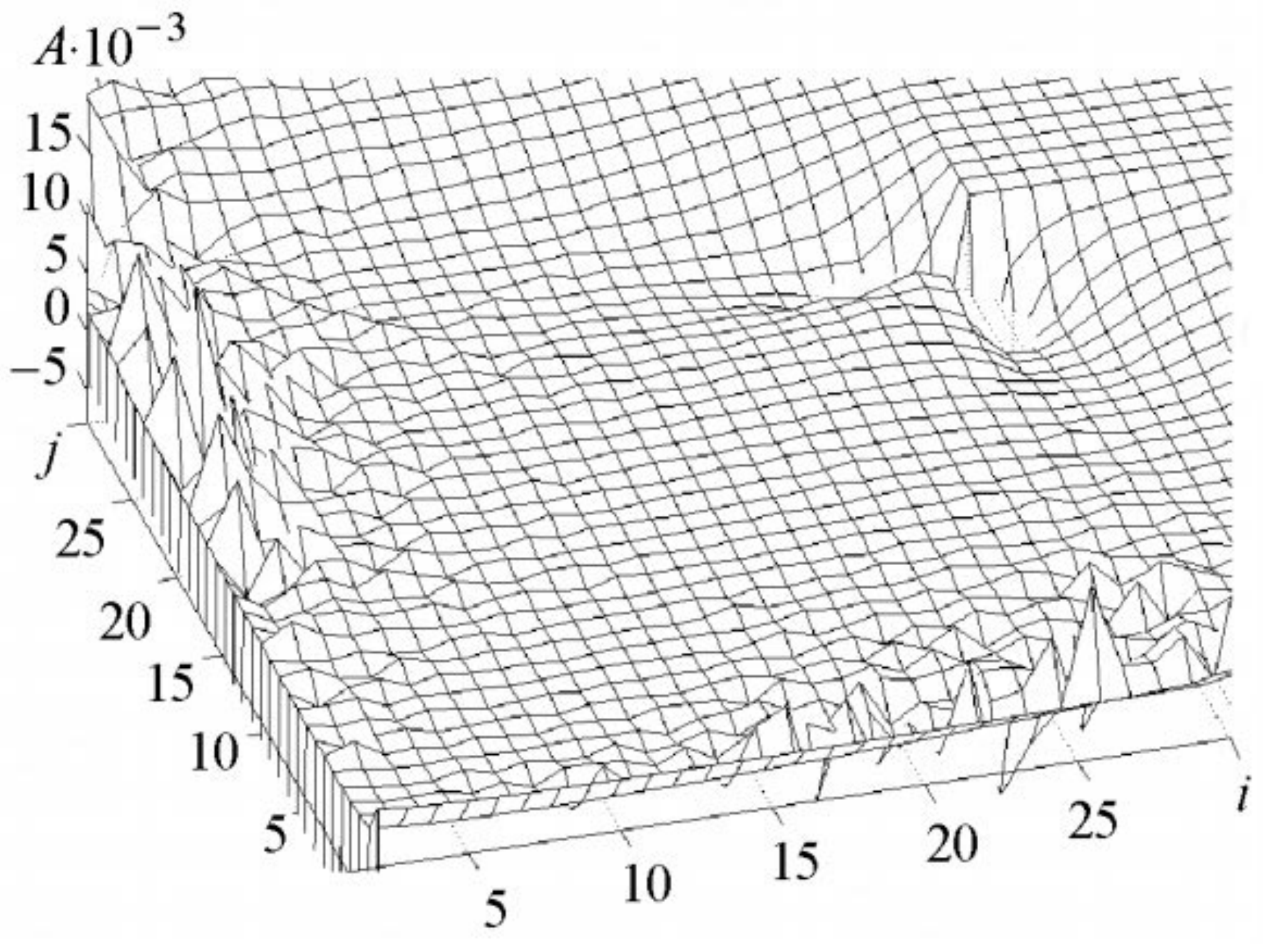}
                 \subcaption{Distribution of \textit{A}.}
                \label{fig:A}
        \end{subfigure}%
        ~~~ 
        \begin{subfigure}[b]{0.45\textwidth}
                \includegraphics[width=1\textwidth]{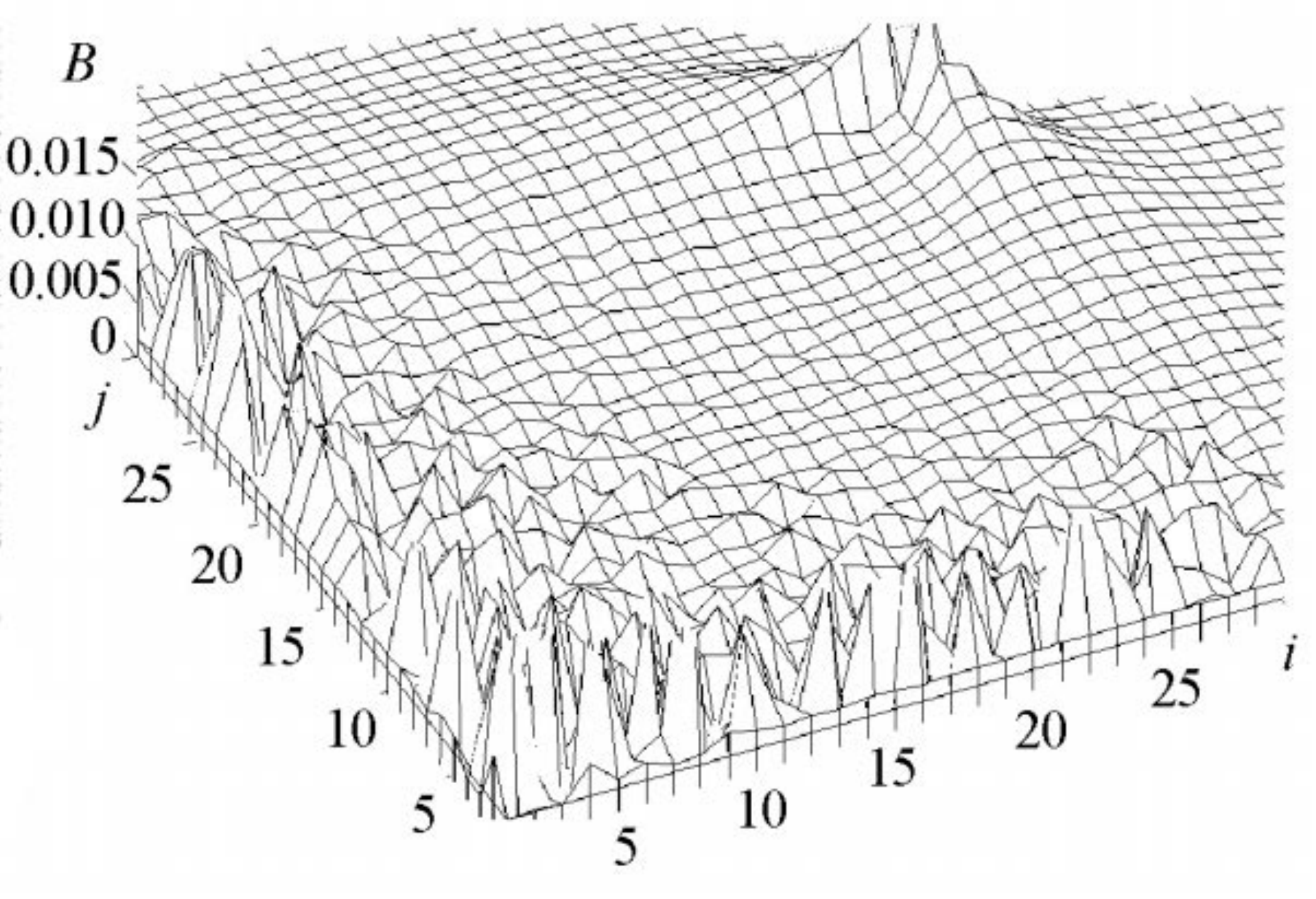}
                  \subcaption{Distribution of \textit{B}.}
                \label{fig:B}
        \end{subfigure}
      \caption{Dependencies of the coefficients in formulae (\ref{counter.u.symmetric}) and (\ref{counter.v.symmetric}) over the computational mesh for the numerical solution to the counterflows
 in a steady regime ($t=30$) at \Rey = 0.1 and \Wi = 4. The minimum step on $x,\ y$ is 0.03.}
           \label{fig:AB}
\end{figure*}

Provide more evidence  of  the counterflows regularity in the central point. It is time now to look at other flow parameters, the stresses and pressure, in comparison with the asymptotic expressions of the previous section.

For that we can determine these expressions' parameters, \textit{A} and \textit{B}, on the basis of the global numerical solution  (cf. \cite{MackarovBif}). This bout, make use of a one-quadrant version of the counterflows solution whose layout is illustrated by Fig.~\ref{fig:1q-lay} and which was shown \cite{Mackarov64} to give the flows patterns closely corresponding  to the full-domain symmetric ones. Overall, the closeness of the solutions forms  can be seen from comparison of Figs.~\ref{fig:StressAnalNum:quarter}, \ref{fig:StressAnalNum} and \ref{fig:PressureAnalNum:quarter}, \ref{fig:PressureAnalNum}, respectively\footnote{To make the comparison as clear as possible, we had to use different scales in the vertical directions of the pairs of figures. Also the axes in the full-domain figures are skipped.}.

It is fairly easy to obtain expressions for \textit{A, B} from a finite-difference form of Eqs. (\ref{counter.u.symmetric}), (\ref{counter.v.symmetric}) by means of the velocities  adjacent nodes values  and asymmetric finite differences:
\begin{equation}
\label{eqn:Ax}
A = N_x/D_x\ ,
\end{equation}
where
\begin{align*}
{N_x} = &3\left( {\left( {{u_{i,j}} - {u_{i + 1,j}}} \right)x_{ij} + \Delta x \cdot {u_{i,j}}} \right){{x_{i,j}}^2}\nonumber\\
& + 3\Delta x \cdot {u_{i,j}}\left( {{{x_{ij}}^2} + x_{ij} \cdot \Delta x + \frac{1}{3}{{(\Delta x)}^2}} \right)\nonumber\\
& + \left( {{u_{i,j}} - {u_{i + 1,j}}} \right){{x_{ij}}^3},
\end{align*}
\begin{equation*}
{D_x} = {x_{ij}} \cdot \Delta x\left( {2{x_{ij}}^2 + 3{x_{ij}} \cdot \Delta x + {{(\Delta x)}^2}} \right),
\end{equation*}
\begin{equation*}
\Delta x = {x_{i + 1j}} - {x_{ij}}.
\end{equation*}
\vspace{0.1cm}
\begin{equation}
\label{eqn:Bx}
B =  - \frac{{3\left( { - {x_{ij}} \cdot ({u_{i + 1,j}} - {u_{i,j}}) + \Delta x \cdot {u_{i,j}}} \right)}}{{{x_{ij}} \cdot \Delta x \cdot \left( {2{x_{ij}}^2 + 3{x_{ij}} \cdot \Delta x + {{(\Delta x)}^2}} \right)}}.
\end{equation}

Similar is another pair of expressions for \textit{A, B} through the mesh values of \textit{v} and \textit{y}:

\begin{equation}
\label{eqn:Ay}
A = N_y/D_y
\end{equation}
where
\begin{align*}
{N_y} = &3\left( {\left( {{v_{i,j}} - {v_{i,j + 1}}} \right)y_{ij} + \Delta y \cdot {v_{i,j}}} \right){{y_{i,j}}^2}\nonumber\\
& + 3\Delta y \cdot {v_{i,j}}\left( {{{y_{ij}}^2} + y_{ij} \cdot \Delta y + \frac{1}{3}{{(\Delta y)}^2}} \right)\nonumber\\
& + \left( {{v_{i,j}} - {v_{i + 1,j}}} \right){{y_{ij}}^3},
\end{align*}
\begin{equation*}
{D_y} = {y_{ij}} \cdot \Delta y\left( {2{y_{ij}}^2 + 3{y_{ij}} \cdot \Delta y + {{(\Delta y)}^2}} \right),
\end{equation*}
\begin{equation*}
\Delta y = {y_{i j+1}} - {y_{ij}}.
\end{equation*}
\vspace{0.1cm}
\begin{equation}
\label{eqn:By}
B =  - \frac{{3\left( { - {y_{ij}} \cdot ({v_{i ,j+ 1}} - {v_{i,j}}) + \Delta y \cdot {v_{i,j}}} \right)}}{{{y_{ij}} \cdot \Delta y \cdot \left( {2{y_{ij}}^2 + 3{y_{ij}} \cdot \Delta y + {{(\Delta y)}^2}} \right)}}.
\end{equation}

\noindent The formulae are valid for both uniform and flexible computational meshes.

As can be seen, the nominators and denominators of the expressions for \emph{A} and \emph{B} contain  $x,\ y$ along with $\Delta x, \Delta y$  as factors. This may result in division of two small numbers and lead to inaccurate results in the vicinity of the axes or on especially fine meshes.

 Meanwhile, Fig.~\ref{fig:AB} presenting \emph{A} and \emph{B} as half-sums of expressions (\ref{eqn:Ax}), (\ref{eqn:Ay}) and (\ref{eqn:Bx}), (\ref{eqn:By}), respectively, shows that their distributions are close to plateaus, i.e., distinct regions of nearly constant values of these coefficients in the neighborhood of the stationary point (i. e., inside the limits of the square with the ''central point -- walls corner`` diagonal, see Fig.~\ref{fig:layout}). Some numerical noise can only be noted nearby the axes. The \emph{A} plateau looks especially smooth --- probably because \textit{A}, compared to \emph{B},  corresponds to a lower power of $x,\ y$ in expansions (\ref{counter.u.symmetric}), (\ref{counter.v.symmetric}) and can be calculated with a smaller error.

 Thus, Fig.~\ref{fig:AB} makes us confident that the asymptotic analytical and global numerical solutions to the counterflows problem indeed well correlate!

In order to increase this confidence, make a more quantitative comparison using the \emph{A} and \emph{B} plateau values $A\approx-0.00611,\ B\approx0.0032$. We get from Eq.~(\ref{counter.Sigma.x}) \ \  $\Sigma_x=-0.0573$,  whereas the numerical solution gives  $\sigma_{xx}=-0.0518$ for the central point --- a good agreement within the involved accuracy if we take into account the above-mentioned peculiarities of formulae  (\ref{eqn:Ax})--(\ref{eqn:By}).

Expressions (\ref{alpha}), (\ref{beta}) give $\alpha = 0.0305,\ \beta = 0.0279$. Closeness of the values suggests symmetry of the normal stress $\sigma_{xx}$ distribution near the stationary point, the one of absolute minimum of the stress. This well corresponds to the numerical solution as clearly seen in Fig.~\ref{fig:StressAnalNum:quarter}.

Finally, we are going to analyze the asymptotic distribution of the pressure. From Eqs.~(\ref{counter.p.x}), (\ref{counter.p.y}) obtain  $P_x = 0.0321$ and $P_y= -0.0321$. Thus, evidently the radius of the pressure profile  curvature  when surfing on it from the stationary point along the \textit{x} direction, and the radius for the case of the \textit{y} direction are equal, with the convexities oppositely directed. This perfectly conforms to Fig.~\ref{fig:PressureAnalNum:quarter}~!

\subsection{Convergence of the numerical solution}
\label{sec:counter:convergence}

\begin{figure*}[t]
  \centering
  \includegraphics[width=0.7\textwidth]{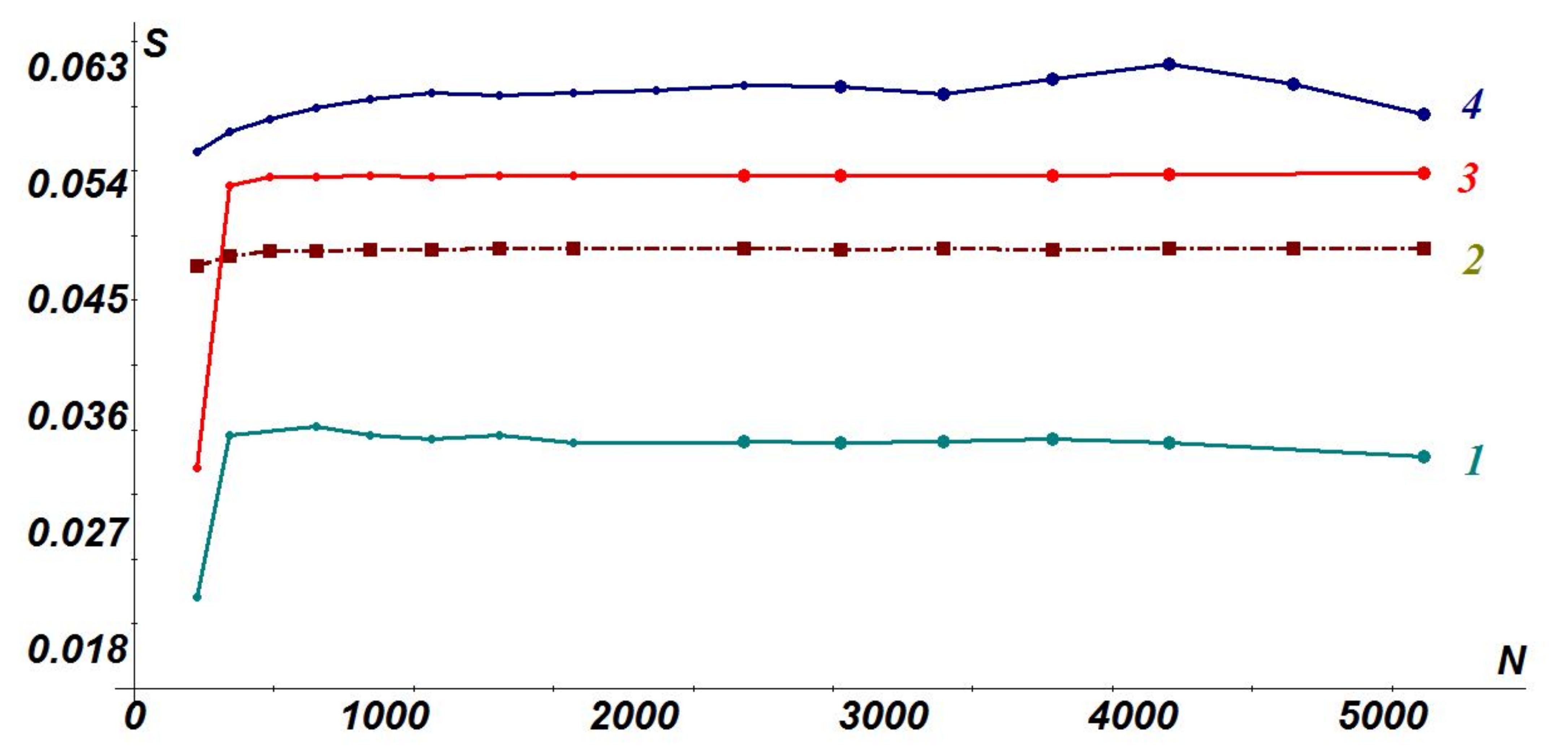}
  \caption{{\footnotesize Sequences of the normal stress $\sigma_{xx}$ absolute values for the numerical solutions to the counterflows  in the stagnation  point --- versus the total numbers of meshes nodes. The stress curves correspond to the following pairs of $\Rey$--$\Wi$ (in the order of their numbering): 0.003--150, 1--1, 0.05--4, 0.01--100. Curves 1,~3,~4 conform to $t=30$ (quasi-stationary regime). Their smaller markers are for  the time step 0.0001; the bigger ones correspond to the time  step 0.00005. Curve 2 is for $t=10$ and $\Delta t = 0.00003$. Minimum steps on $x,\ y$ for each of the horizontal axis total nodes numbers are 0.109,~ 0.0771,~ 0.0630, 0.0545, 0.0488. }}
  \label{fig:FullDomainConv}
\end{figure*}

As to the full-domain problem statement (Fig.~\ref{fig:layout}), it was possible to obtain stable solutions with \textit{Wi} up to 150 and with \textit{Re} such that $\Wi \cdot \Rey \lesssim 1$ (cf. \cite{Hulsen} with \textit{Wi}=100, \textit{Re}=0.01).

Let us demonstrate the convergence of the full-domain solutions following \cite{Mackarov2011}, where  the one-quadrant case was considered in detail.

Usually one proves convergence by a comparison of computational data obtained with various finite difference steps on  independent variables (different numbers of mesh nodes).

Since the mesh used in calculations is flexible, probably the simplest and most evident way to demonstrate the used numerical procedure convergence and reliability is to examine the value of normal stress, for example, $\sigma_{xx}$ in one of the most critical and interesting points of the solution, \textit{the central stagnation point}. The distribution of $\sigma_{xx}$ round there has a sharp minimum (Figs.~\ref{fig:StressAnalNum:quarter}, \ref{fig:StressAnalNum}), large space derivatives and may have especially high error of approximation. Evidently, as long as the sequence of these values converges with the mesh getting finer, one can expect still better convergence in the rest of the domain.

As seen in Fig.~\ref{fig:FullDomainConv}, with finer meshes the time step was chosen smaller, so as to satisfy the condition of the pressure correction method convergence (see Section \ref{sec:NumDetails} and the Appendix).

Aside stays the dashed curve 2 ($\Wi=\Rey = 1$). Here an especially small time step was needed to get a solution on all the meshes. Yet, the solution sharply exploded before the reach of the stationary regime, by the moment $t\approxeq15$. This fact does not harm, however, the look and feel of the solution's stable behavior before the crash. The described picture is typical for the obtained solutions with a relatively small viscosity and most probably points to an inertial instability, which is really hard to deal with numerically.

As to higher Weissenberg numbers, the results reveal certainly good convergence of the stagnation point stress values. The data points form an ideal plateau for $\Wi=4$. The plateaus 1, 4 of really high $\Wi\ge 100$ are not so perfect. Evidently this is because
\begin{itemize}
\item  a flow in this case is very dynamic and non-stationary (periodic emergence of vortexes never vanishes),
\item  the calculation time interval is comparatively large: $0 \le t \le 30$,
\item the number of time steps is essential.
\end{itemize}
So, a lot of calculations was needed to reach the final moment. That conceivably  resulted in a noticeable accumulation of computational errors.

To get convinced in the solution quality, let us look at Figs.~\ref{fig:VortexCounterPattern}, \ref{fig:RegularCounterPattern}, \ref{fig:StressAnalNum}, \ref{fig:PressureAnalNum}, \ref{fig:FullDomainConv}  altogether. The figures suggest that on the whole set of the used meshes the solution is sustainable, smooth and preserves its salient features, above all, the counterflows' periodic reversals accompanied by emergence of vortexes. 

\paragraph{Corners.} What is also seen in Fig.~\ref{fig:PressureStressAnalNum} is that pressure and stress gradients near the slots corners get very high. In fact, they are especially high with  $\Rey \lesssim 1$. Interestingly, the pressure  and stress gradients herewith nearly balance each other so that the velocity field is smooth and satisfies the no-slip conditions. A thorough look at the solutions reveals that such peculiarities at the corners are mostly local, converge to finite values, and do not remarkably affect the flow as a whole and its part near the stagnation point.

\section{Flow spread over a wall}
\label{sec:spread}

\begin{figure*}[t]
	\centering
	\begin{subfigure}[b]{0.1485\textwidth}
		\includegraphics[width=1\textwidth]{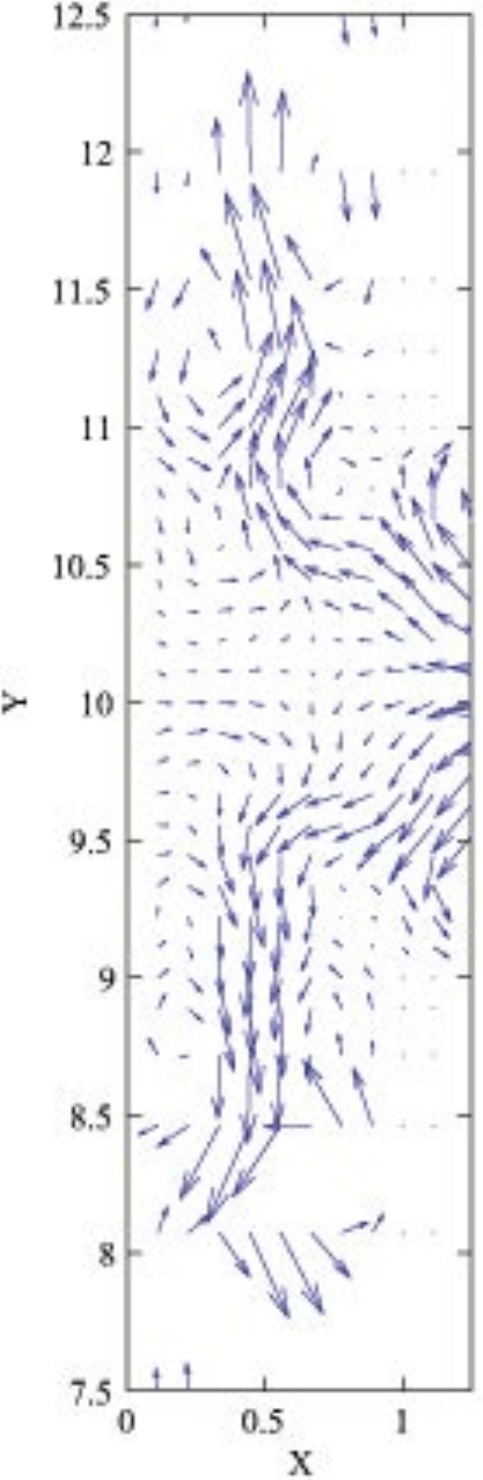}
		\caption{\scriptsize{$\Delta t = 5\cdot10^{-4}$,\\ $ \Delta = 0.10$.}}
	\end{subfigure} ~~
	\begin{subfigure}[b]{0.1515\textwidth}
		\includegraphics[width=1\textwidth]{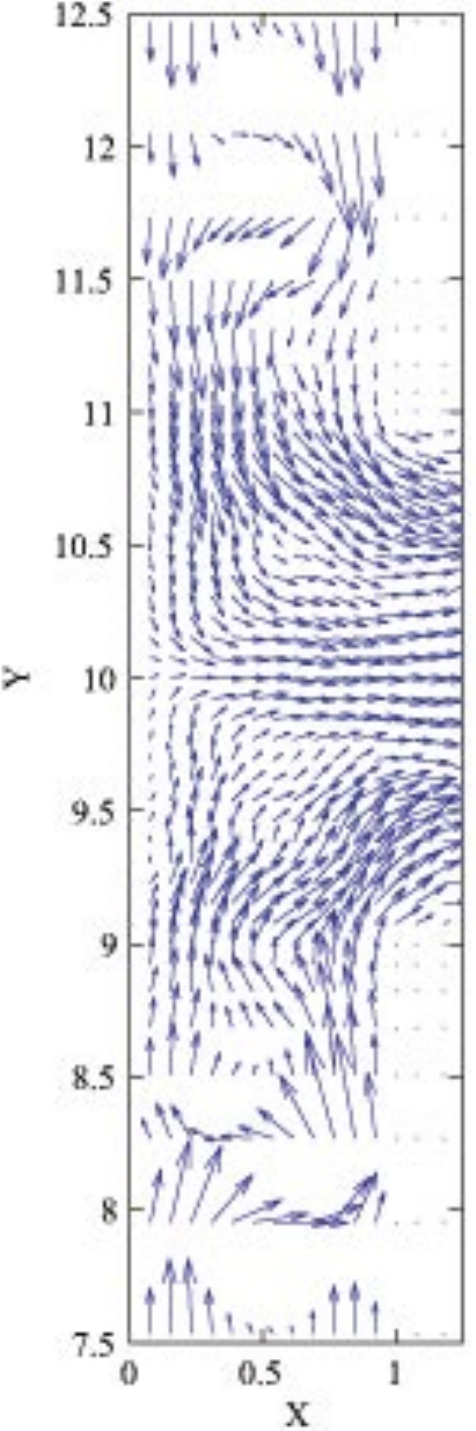}
		\caption{\scriptsize{$\Delta t =5\cdot{10^{-4}}$,\\  $\Delta = 0.071$.}}
	\end{subfigure} ~~
	\begin{subfigure}[b]{0.1505\textwidth}
		\includegraphics[width=1\textwidth]{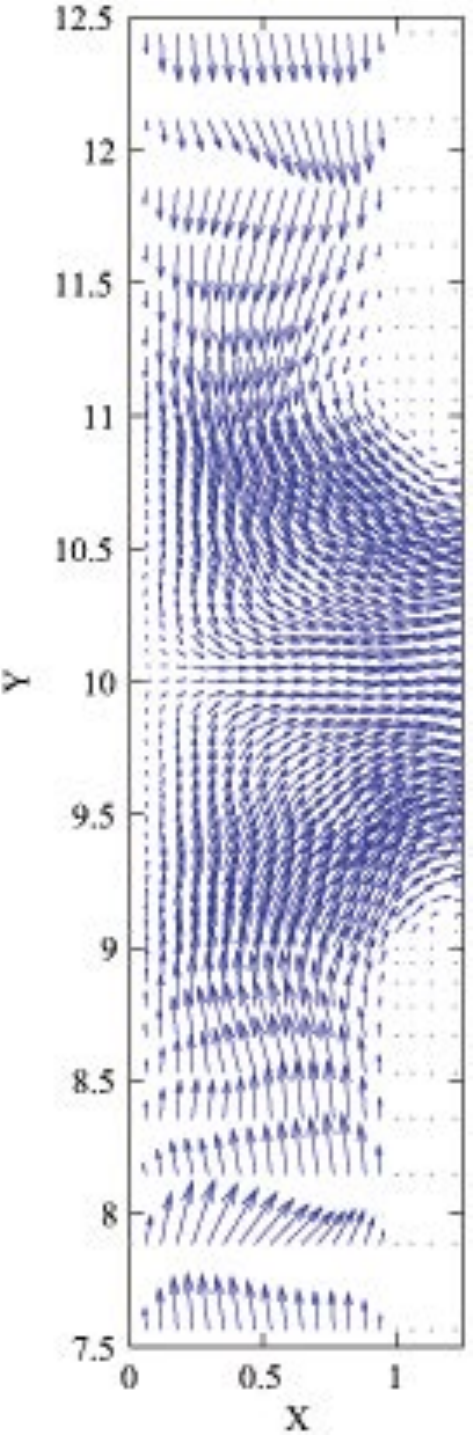}
		\caption{\scriptsize{$\Delta t =10^{-4}$,\\  $\Delta = 0.056$.}}
	\end{subfigure} ~~
	\begin{subfigure}[b]{0.1495\textwidth}
		\includegraphics[width=1\textwidth]{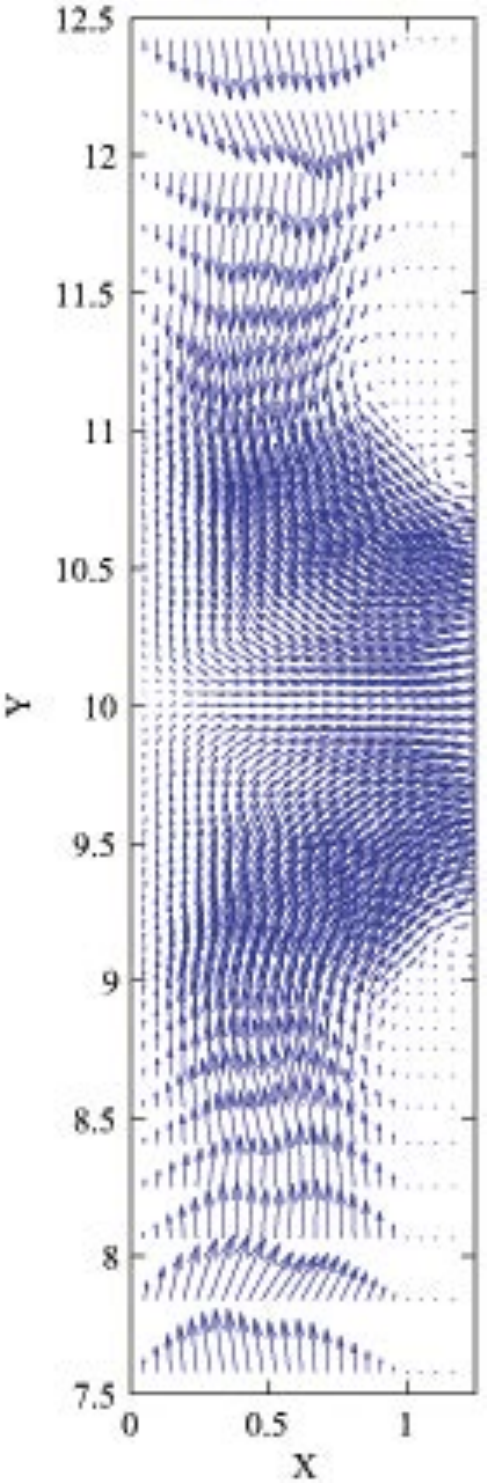}
		\caption{\scriptsize{$\Delta t =10^{-4}$, \\  $\Delta = 0.046$.}}
	\end{subfigure}
	\caption{An instance of the spread over a wall  given by the numerical simulation on different meshes; $t=4.045$, $\Rey=0.003$, $\Wi=150$. Specified are the steps on time and mesh: $\Delta t$ and $ \Delta$.}
	\label{fig:VortexSPreadPattern}
\end{figure*}

Since the subject matter of this paper concerns the whole class of the UCM fluid benchmark flows, shortly speak of another its representative, the flow along a horizontal slot reaching a vertical wall and spreading over it. Figure~\ref{fig:VortexSPreadPattern} shows a snapshot of such a flow, with the pressure inlet and outlet boundary conditions the same as in the case of counterflows. Similarly to this case,  we observe a \textit{regular} pattern, and vortex-like structures accompanying the flow reversal.

For the sake of a reasonable volume of this paper, we are going to put up the detailed presentation of this flow --- with providing an asymptotic analytical solution, its comparison with the numerical simulation, analysis of numerical convergence etc. --- for a new article.

For now mention, however,  the statement of \cite{Becherer,Objections} in regard with the elongational and wall stagnation flows: these flows' stationary points definitely correspond to singularities in the solutions to the UCM constitutive state equation... The asymptotic analytical solution of ours does not confirm this statement \emph{with respect to the wall stagnation flow}.  As seen in Fig.
~\ref{fig:VortexSPreadPattern}, the solution is regular and numerically satisfies the problem governing equations (including the UCM state equation) together with the boundary/symmetry conditions.

 \section{Details of the numerical procedure}
 \label{sec:NumDetails}

 \paragraph{Mesh.}
Both the counterflows and spread layouts (Figs.~\ref{fig:layout}, ~\ref{fig:VortexSPreadPattern}) involve a quadrangle between the walls corners. It is located between the other parts of the computational domain, horizontal and vertical slots (four or three, respectively). All five or four parts have equal numbers of steps on \textit{x}, \textit{y}. The steps are hereby uniform within the central quadrangle, and  flexibly (in geometrical progression) diminishing towards the quadrangle (and towards the value of a step inside it).

 \paragraph{Computational scheme.}
In the evolutionary equations an explicit first-order finite-difference approximation of the time derivatives was used. In the momentum equation the pressure gradient was approximated by the backward finite differences. In the UCM state equation the spatial derivatives of the stresses and velocities were described by the forward finite differences, while for the derivatives of the stresses in the momentum equation  the backward differences were again used.  The convective terms in the momentum equation were represented by the upwind finite differences.

Such a choice of finite differences was dictated, \textit{inter alia}, by usage of the pressure correction method.

\paragraph{Pressure correction method.}
The applied numerical procedure is essentially based upon a simple and pretty efficient
version of the pressure correction method (hereafter also PCM). With the details of the method itself available in the Appendix, discuss here some features and the overall value of the method in this work.

At each time step the iterative procedure of PCM adjusts the pressure and velocity so as to satisfy the continuity equation and thus reach the incompressible fluid state. The iterations are terminated when the norm of the velocity divergence (defined in the Appendix) reaches the value of $10^{-5}$, which is much smaller than the smallest step on a space variable for any of the meshes used in practice. Thus the error brought by PCM is negligent compared to the space approximation errors.

It was practically detected that in nearly all the kinds of calculations the fastest PCM convergence was achieved with its parameter $\mu \approx 0.7$ (see Eq.~(\ref{pij})).

Herewith in the case of the counterflows it usually took PCM 1--2 iterations to correct the pressure on stable flow phases, whereas on top of the vortexes' intensities 10--12 ones were often needed.

As to the spread flow, much more iterations were sometimes required: up to 200-300 for $t<1$, when the flow was the least stationary. A slower convergence conceivably follows from a specific, more fine-grained structure of this kind of a flow.

In relation with the hitherto stated opinions about singularities as a characteristic feature of the UCM model \textit{per se} or poor compliance between the UCM rheological law and momentum equation, a word must be said about a possible role of PCM in obtaining regular solutions in this research. At any PCM iteration the corrected pressure and velocity values in every node depend upon the pressure and velocity in its neighboring nodes, which in their turn are affected by the joint nodes values, and so forth...  Thus, one can say about some \textit{integral} nature of PCM, its ability to bring \textit{hyperbolic features} to the numerical set of equations. It seems to be able to ``smear out'' sharp short-wave disturbances (once they evidently can  corrupt the incompressibility \textit{locally}) --- specific for high elasticity --- over the whole of the flow domain. It is this feature of PCM that most probably makes the method successful in dealing with high Weissenberg numbers.

This PCM trait is worth deeper and more formal investigation in the future.

\section{Conclusions and discussion}

On the ground of the presented research we thus cannot  answer positively the paper's title question. 

Indeed two typical benchmark flows of the UCM fluid, were investigated, with counterflows thoroughly studied using an analytical and numerical procedures.  Both analytical and numerical approaches were based on rigorous meeting the governing relations, \textit{viz.} the momentum, continuity and UCM constitutive state equations, as well as physically natural boundary conditions. As far as the momentum equation is concerned, its \textit{convective terms} were always kept, since even for high viscosities they were essential at least near stagnation points and inside vortexes because of high strains.

 So there was no need to overburden the problems statement by additional pre-study suppositions (for example,  about a flow velocity field or distributions of the stresses over the space \cite{VanGorder}). As a result, clearly regular behavior of the flows in the region of stationary points was observed.

Thus, singular stresses are not believed to be an intrinsic feature of the UCM rheological model. This does not mean, however, that it never involves sharp peaks of stresses. Such are the distinct stresses extrema observed in both kinds of the flows at the walls corners.

Though this special  behavior of the  stresses does not seem to affect the regularity of a flow near a stagnation point, a flow of a viscoelastic fluid near a corner as such is an interesting subject matter for future investigations. Especially attractive would be finding an analytical solution to the flow in a region like this. No doubt, such a research would generally deepen the comprehension of viscoelasticity.

 \section*{Appendix. Convergence of the pressure correction method}

The numerical procedure employed in the present research involves PCM in the version having conceivably higher performance than the ones available in literature \cite{Anderson, Ferziger}. This is, in particular, due to the mesh having a simple topology, and since there is no need to solve the Laplace equation at each time step. For the sake of completeness, provide a formal proof of convergence of the method in such a version, some more detailed compared to \cite{Mackarov2009,MackarovMZG}.

Consider a two-dimensional non-stationary flow of an incompressible fluid \textit{without any specific suppositions about its rheological nature}. So, the procedure involved can be  applied to other rheological laws besides the UCM model. In particular, this version of PCM was earlier successfully used for simulation of the Newtonian fluid flows (\textit{vide} \cite{Mackarov2009}).

Making use of an explicit first-order approximation of the time derivatives  in the components of the momentum equation (\ref{momentum}), rewrite them in a finite-difference form involving \textit{backward difference operators for the pressure} and linear operators in a common appropriate form for the rest of the terms:
\begin{equation}
u_{ij}^{n + 1} = u_{ij}^n + ({M_u}(u_{ij}^n,\ v_{ij}^n,\ \sigma _{11\;ij}^n,\ \sigma _{12\;ij}^n) - {\Delta _{x - }}{p_{ij}}) \cdot \Delta t,
\label{uij}
\end{equation}
\begin{equation}
v_{ij}^{n + 1} = v_{ij}^n + ({M_v}(u_{ij}^n,\ v_{ij}^n,\ \sigma _{12\;ij}^n,\ \sigma _{22\;ij}^n) - {\Delta _{y - }}{p_{ij}}) \cdot \Delta t.
\label{vij}
\end{equation}
Given the velocity field on the $n^{th}$ time layer  satisfying the continuity equation (\ref{continuity}), the new field  given by Eqs.~(\ref{uij}), (\ref{vij})  does not generally satisfy it, so that an adjustment is needed. For that specify pressure on the new time layer as follows:
\begin{equation}
p{_{ij}^{n + 1\;{\kern 2pt}\prime}} = p_{ij}^{n + 1} - \mu \,div\,\bar v_{ij}^{n + 1}
= p_{ij}^{n + 1} - \mu \,({\Delta _{x + }}u_{ij}^{n + 1} + {\Delta _{y + }}v_{ij}^{n + 1}),\quad \mu  > 0
\label{pij}
\end{equation}
meaning herewith
\begin{equation*}
p_{ij}^{n + 1} \equiv p_{ij}^n\,\,\,\,\,\forall \,i,j.
\end{equation*}
Again determine the velocities by Eqs.~(\ref{uij}), (\ref{vij}) and correct the pressure via Eq.~(\ref{pij}) to obtain the following iterative procedure (henceforth instead of indexes \textit{i } and \textit{j},\  \  $\tau$, the number of an iteration, will be used):
\begin{equation}
{p_{_{\tau  + 1}}} = {p_{_\tau }} - \mu \;div  \,{\bar v_\tau },
\label{ptau}
\end{equation}
\begin{equation}
u_{\tau  + 1}^{n + 1} = u_\tau ^{n + 1} - \Delta t{\kern 1pt} \;({\Delta _{x - }}{p_{\tau  + 1}} - {\Delta _{x - }}{p_\tau }),
\label{utau}
\end{equation}
\begin{align}
v_{\tau  + 1}^{n + 1} = v_\tau ^{n + 1} - \Delta t{\kern 1pt} \;({\Delta _{y - }}{p_{\tau  + 1}} - {\Delta _{y - }}{p_\tau }).
\label{vtau}
\end{align}

\emph{As long as the procedure converges}, the divergence in Eq.(\ref{ptau}) tends to zero so the velocities and pressure for the next time layer are close to those of an incompressible fluid.

Let us then prove the convergence of iterative procedure (\ref{ptau})-(\ref{vtau}) to the incompressible state.

Applying finite right-side differentiation operators ${\Delta _{x + }}$ and ${\Delta _{y + }}$ to Eqs. (\ref{utau}) and (\ref{vtau}), respectively, and summing them with the use of Eq.~(\ref{ptau}), we get an equation for the velocity divergence:
\begin{equation}
\frac{{div\,\bar v{{_{\tau  + 1}^{n + 1}}} - div\,\bar v{{_\tau ^{n + 1}}}}}{{\Delta   t}} 
= \mu   {\kern 1pt} ({\Delta _{xx}}div\,\bar v{_\tau ^{n + 1}} + {\Delta _{yy}}div\,\bar v{_\tau ^{n + 1}}).
\label{thermo}
\end{equation}
Analyzing this equation, we will be assuming that the velocity divergence is a doubly continuously differentiable function of the spatial variables. Designate it \emph{d}. In Eq.(\ref{thermo}) $\Delta_{xx}$ (and accordingly $\Delta_{yy}$) is expressed as follows:
\begin{equation*}
{\Delta _{xx}}{d_{ij}} = \frac{{{d_{i + 1{\kern 1pt} j}}({x_i} - {x_{i - 1}}) - {d_{ij}}({x_{i + 1}} - {x_{i - 1}}) + {d_{i - 1{\kern 1pt} j}}({x_{i + 1}} - {x_i})}}{{{{({x_{i + 1}} - {x_i})}^2}({x_i} - {x_{i - 1}})}}
\end{equation*}
getting simplified in the uniform mesh case:
\begin{equation*}
{\Delta _{xx}}{d_{ij}} = \frac{{{d_{i + 1{\kern 1pt} j}} - 2{\kern 1pt} {d_{ij}} + {d_{i - 1{\kern 1pt} j}}}}{{\Delta x}^2}.
\end{equation*}
The right-hand side of Eq.~(\ref{thermo}) can be presented, therefore, as a product of a tridiagonal matrix \textit{M} with ``non-strict diagonal dominance'':
\begin{align}
\left| {{M_{kk}}} \right| = \sum\limits_{k \ne l} {\left| {{M_{kl}}} \right|} ,\quad 1 \le k,l \le K
\label{3M}
\end{align}
... and a vector of the divergence mesh values. This is true for both uniform and non-uniform meshes. Note that in this expression the $k, l$ subscripts, unlike $i, j$, are the \textit{absolute} numbers of the mesh nodes up to the maximum number $K$.

Then, assuming $\Delta t < 1/\mu {\kern 1pt} {M_{kk}}$  (this poses a limitation on the time step to guarantee the convergence) in every mesh node we have an estimate  for two differences between the consecutive iterations:
\begin{equation}
\left| {d_k^{\tau  + 2} - d_k^{\tau  + 1}} \right|
\le \left( {1 - \mu {\kern 1pt} \Delta t{\kern 1pt} {M_{kk}}} \right)\left| {d_k^{\tau  + 1} - d_k^\tau } \right| 
+ \mu {\kern 1pt} \Delta t\sum\limits_{\;l \ne k} {{M_{kl}}\left| {d_l^{\tau  + 1} - d_l^\tau } \right|}.
\label{consecutive}
\end{equation}
Eq.~(\ref{ptau}) suggests that specifying a constant pressure at the inlet/outlet leads to a condition for the divergence
\begin{equation}
{d_{inlet/outlet}} = 0
\label{dio}
\end{equation}
from which it can be easily seen that not all the differences of the divergence mesh values in inequality~(\ref{consecutive})  are of the same sign. This makes it possible to replace the weak inequality sign there by the strict one. Further, replace the  absolute values by a norm defined as $\left\| d \right\| \equiv \mathop {\max }\limits_{1 < k < K} \left| {{d_k}} \right|$. In relation with Eq.~(\ref{3M}) this brings about
\begin{equation}
\left\| {d_{}^{\tau  + 2} - d_{}^{\tau  + 1}} \right\| < \left\| {d_{}^{\tau  + 1} - d_{}^\tau } \right\|,
\label{inequality}
\end{equation}
which proves  the convergence of the iterative procedure (\ref{thermo}) to the identical zero solution. The procedure (\ref{ptau})--(\ref{vtau}) therefore converges to the pressure and velocity corresponding to the incompressible fluid state!

\subparagraph{On a variety of boundary conditions.}

The boundary conditions for the flows of this paper are also pertinent to a great many other important problems of fluid dynamics. In particular, to those with a combination of the \textit{inlet/outlet pressure} and \textit{no-slip} conditions.

The \textit{former} condition, in terms of the velocity divergence, was shown above to assure the PCM convergence.

As to the \textit{latter}, it is equivalent to a second-order boundary condition: zero value of the divergence wall normal derivative. To make sure of this,  one only needs to look at Eqs. (\ref{utau}), (\ref{vtau}). For a wall mesh node, with zero velocity, they will give
\begin{equation*}
{\Delta _{x - }}{p_{\tau  + 1}} - {\Delta _{x - }}{p_\tau } = {\Delta _{y - }}{p_{\tau  + 1}} - {\Delta _{y - }}{p_\tau } = 0
\end{equation*}
and consequently
\begin{equation*}
{\Delta _{x - }}{d_\tau } = {\Delta _{y - }}{d_\tau } = 0\,\,\,\,\,\,\forall \tau
\end{equation*}
for a wall node\footnote{Without loss of generality assume the boundary to be above and on the right of the domain. Otherwise, we can change the differences direction.}, which actually leads to
\begin{equation}
{\left( {{\raise0.7ex\hbox{${\delta d}$} \!\mathord{\left/
				{\vphantom {{\delta d} {\delta n}}}\right.\kern-\nulldelimiterspace}
			\!\lower0.7ex\hbox{${\delta n}$}}} \right)_{wall}} = 0.
\label{normal}
\end{equation}

So, for the pressure correction process to be compliant with the inlet/outlet and no-slip boundary conditions, Eqs. (\ref{dio}) and (\ref{normal}) must be satisfied. Fortunately, both these constraints comply with the identical zero solution for the velocity divergence.

The actual version of PCM promises to be effective with other practical forms of boundary conditions too. For instance, suppose that, instead of pressure, \textit{ volume fluxes} of the fluid are specified at the inlet/outlet. Then we still have the strict form of inequality (\ref{consecutive}) following from the invariance of the stationary flow domain (rigid walls), and the fluid incompressibility, which requires
\begin{align*}
\sum\limits_{k = 1}^K {{d_k}}  = 0
\end{align*}

It also seems possible to involve certain slip conditions, as any physically reasonable slip law is believed to preserve diagonal dominance of the matrix from relation (\ref{consecutive}).

These considerations may advise this form of PCM as a valuable means for solving various problems of fluid mechanics.

\bibliographystyle{plain}
\bibliography{Mackarov}
\end{document}